\documentclass{aa}  

\usepackage{graphicx}
\usepackage{txfonts}
\usepackage{graphicx}
\usepackage{lmodern}
\usepackage{float}
\usepackage{caption}
\usepackage{comment}
\usepackage{amsmath,amsfonts,amssymb,mathrsfs,xfrac}
\usepackage[colorlinks=True,citecolor=blue,linkcolor=blue]{hyperref}
\usepackage[normalem]{ulem}

\usepackage{xcolor}

\newcommand{\AM}[1]{\textcolor{black}{#1}}

\renewcommand{\vec}[1]{\boldsymbol{#1}}
\newcommand{\sepa}{s}
\newcommand{\kthree}{\vec{\underline{k}}}

\newcommand{\xthree}{\vec{\underline{x}}}

\newcommand{\dd}{{\rm d}}

\newcommand{\fovarea}{\mathcal{S}_{\mathcal{A}}}
\newcommand{\meansf}{\langle SF \rangle}

\newcommand{\ww}{\mathcal{W}}
\newcommand{\wwp}{\mathcal{W}^{\prime}}

\begin{document} 

\title{Toward mapping turbulence in the intracluster medium }
\subtitle{III. Constraints on the turbulent power spectrum with Athena/X-IFU}
\titlerunning{Toward mapping turbulence in the intra-cluster medium. }

\author{S. Beaumont\inst{1,2},
        A. Molin\inst{1}\thanks{Corresponding author : A. Molin - alexei.molin@irap.omp.eu},
        N. Clerc\inst{1},
        E. Pointecouteau\inst{1},
        M. Vanel,
        E. Cucchetti \inst{2},
        P. Peille \inst{2},
        F. Pajot \inst{1}
        }

\authorrunning{S. Beaumont, A. Molin}

\institute{IRAP, Université de Toulouse, CNRS, CNES, UT3-PS, Av. du Colonel Roche 9, 31400, Toulouse, France
             \and
    Centre National d’Etudes Spatiales, Centre spatial de Toulouse, 18 avenue Edouard Belin, 31401 Toulouse Cedex 9, France
           }
\date{Received ...}

\abstract
{Future X-ray observatories with high spectral resolution and imaging capabilities will enable measurements and mappings of emission line shifts in the intracluster medium (ICM). Such direct measurements can serve as unique probes of turbulent motions in the ICM. Determining the level and scales of turbulence will improve our understanding of the galaxy cluster dynamical evolution and assembly, together with a more precise evaluation of the non thermal support pressure budget. This will allow for more accurate constraints to be placed on the masses of galaxy clusters, among other potential benfits.}
{In this view, we implemented the methods presented in the previous instalments of our work to characterising the turbulence in the intra-cluster medium in a feasibility study with the X-ray Integral Field Unit (X-IFU) on board the future European X-ray observatory, Athena.}
{From idealized mock observations of a toy model cluster, we reconstructed the second-order structure function built with the observed velocity field to constrain the turbulence. We carefully accounted for the various sources of errors to derive the most realistic and comprehensive error budget within the limits of our approach. With prior assumptions on the dissipation scale and power spectrum slope, we constrained the parameters of the turbulent power spectrum model through the use of Markov chain Monte Carlo  (MCMC) sampling. }
{With a very long exposure time, a favourable configuration, and a prior assumption of the dissipation scale, we were able to retrieve the injection scale, velocity dispersion, and power spectrum slope, with 1$\sigma$ uncertainties for better than $\sim$15\% of the input values. We demonstrated the efficiency of our carefully set framework to constrain the turbulence in the ICM from high-resolution X-ray spectroscopic observations, paving the way for more in-depth investigation of the optimal required observing strategy within a more restrictive observational setup with the future Athena/X-IFU instrument.}
{}


\maketitle
%
\section{Introduction}

%
\indent Turbulence is one of the non-gravitational processes ruling the dynamics and thermodynamics of the ionized plasma found in groups and clusters of galaxies, contributing to its virialization and to the subtle balance between cooling and heating processes  \citep[see e.g.,][]{Simionescu2019,Kunz2022}. Various sources can inject turbulence in the intragroup and intracluster medium (ICM hereafter). \\
%
\indent One of the sources is the activity of the super-massive black holes (SMBHs) hosted at the center of galaxies within the cluster. These SMBHs eject large amount of energy via powerful jets. This process, known as AGN feedback, balances the gas cooling at the center of groups and clusters of galaxies, quenches star formation in galaxies, and displaces gas. This generates turbulent motions in massive halos out to scales of hundreds of kpc \citep[see e.g.,][]{McNamara2012,Eckert2021,2022hxga.book....5H}.  \\
%
\indent A second source of turbulence is the accretion and merger of sub-halos, namely, massive galaxies and small groups, falling within the main halo potential well. Depending on their initial dynamics and the accretion and merger configuration, they tend to fall in a spiraling motion around the center of the main halo in sloshing motions that displace gas and induce turbulence \citep[see e.g.,][]{Markevitch2007, ZuHone2022}.  \\
%
\indent A third notable source of turbulence is the growth of groups and clusters via the continuous accretion of dark matter, gas, and galaxies fed by the filamentary structures of the cosmic web to which they are connected. It also happens through merger events with other massive halos. \\
\indent All these processes whether it is feedback, accretion, sloshing, generate shocks at all scales, where the released gravitational energy is transferred to the accreted gas mainly as kinetic energy, heating it to the temperatures of $10^7-10^8$~K. However, a fraction is channeled as non-thermal energy in turbulent motions in the gas which cascade in eddies down to viscous scales, thereby contributing to the gas heating \citep[see e.g.,][]{deGasperin2017, Gaspari2017,Voit2018, Biffi2022}. 
Numerical simulations of structure formation predict that the fraction of non-thermal pressure induced by turbulent motions in the total budget of the intragroup and intracluster gas is 10-30\% \citep{Lau2009, Vazza2009, Nelson2014, Shi2015,Biffi2016}. This makes it a non negligible contribution to the overall gas pressure budget, which has to be accounted for, for instance, in estimation of the halo mass via the hydrostatic equilibrium \citep[see e.g.,][]{Pratt2019}. \\
%
\indent As turbulent motions displace gas, they produce a velocity distribution of the gas particles within the halo referential, inducing a well-known Doppler-shift and broadening of spectral emission \citep{Inogamov2003}. Although emission lines from many chemical elements in the ICM have been observed for decades at X-ray wavelengths \citep[see e.g.,][]{Mernier2018,2022hxga.book...12M}, their proper characterization requires high spectral resolution. It is needed to disentangle the line broadening and centroid shift due to turbulent and gas motions  from other contributors (e.g., thermal broadening, instrument response). To date, this has been achieved only with measurements from gratings instruments, such as  RGS on board XMM-\textit{Newton} \citep{Pinto2020}. However, the lack of spatial resolution and sensitivity of gratings mixes spatial scales and restricts measurement to the very core of the brightest clusters of galaxies. The only actual spectral high resolution measurements of lines centroid and broadening came from the SXS instrument  on board the \textit{Hitomi} satellite, in the core of the Perseus clusters \citep{HitomiCo2016a, HitomiCo2018c}.\\
\indent Despite these instrumental limitations, indirect measurements of turbulence within the ICM have been obtained through various methods, such as the statistics of surface brightness fluctuations. These are assumed to trace the density fluctuations induced by turbulent gas motions \citep{Churazov2012, Zhuravleva2014,Zhuravleva2018,Simionescu2019,Dupourque2023}. Another method relies on the inter-calibration of the Cu-K$\alpha$ line between data from the instrumental background and the observations \citep{Sanders2020,Gatuzz2022}. As initially stressed by \citet{Churazov2012} the interpretation of density fluctuations as due to turbulent motions is one of several, as other processes could produce similar fluctuations (e.g., perturbations of the gravitational potential or entropy fluctuations due to infalling cold gas). \\
\indent As a consequence, direct measurements (i.e., an observable directly linked to the speed of the medium) are needed to characterize the properties of the turbulence in groups and clusters of galaxies. The closest direct measurements available are high-resolution spectra at X-ray wavelengths, which provide the only way to properly measure the associated velocities.
The next generation of X-ray observatories will embark integral field units with the necessary high spectral resolution of the order of a few eV (to be compared to the $\sim$150eV resolution of the currently operating X-ray spectro-imager, that is EPIC on board XMM-\textit{Newton} or ACIS on board \textit{Chandra}). The \textit{Resolve} instrument \citep{Ishisaki2022} on board the XRISM mission \citep{Ueda2022} will soon provide measurements of the ICM emission line shifts and broadenings thanks to its $\sim$5~eV resolution over its 6$\times$6 pixels matrix (over 3$\times$3~arcmin$^2$), albeit with a limited spatial resolution of $\sim$1~arcmin FWHM. In the mid-2030s, the European Space Agency flagship mission \textit{Athena} \citep{Barret2022b}, with its X-ray Integral Field Unit instrument \citep[X-IFU][]{Barret2022}, will bring the spectral resolution to 2-3~eV  and a spatial resolution of 5-10~arcsec. Associated with a large collecting area, providing sufficient sensitivity and a decent field of view ($\sim$5~arcmin diameter), it will allow for a realistic and detailed line-of-sight-integrated velocity mapping of the ICM in groups and clusters of galaxies.

%
\indent To fold such measurements into actual constraints on the properties of the turbulent motion and its associated power spectrum (e.g., velocity dispersion, the injection and dissipation scales) a statistical treatment of the relevant diagnostics, such as the line broadening or centroid shift, is required. Structure functions are related to the two point correlation functions of the projected velocity field underlining the 3D power spectrum of turbulent motions \citep{Inogamov2003, Zhuravleva2012}. They can be used to characterize of the turbulence in the ICM.
\citet{ZuHone2016} used this diagnostic tool to demonstrate the capacity of Hitomi to map turbulence at the center of the Coma cluster. Following these authors, \citet{Roncarelli2018} applied this formalism to \textit{Athena}/X-IFU simple mock observations giving a hint of how \textit{Athena}/X-IFU will allow for breakthrough measurements of the velocity structure induced by bulk motions and turbulence. \\
\indent Moreover, the treatment of the stochastic nature of the turbulent process raises the need to account for the limited number of observed realisation (i.e., one per cluster). This implies an associated sample (or cosmic) variance, that is the variance due to a finite number of available observations and samples, which is a key component to the error budget, although this is difficult to estimate. The aforementioned works partially accounted for this error component, making use of a Monte Carlo approach that involves the estimation of errors from a large number of time consuming mock observations. \\
\indent In our first two papers \citep{Clerc2019,Cucchetti2019}, we have developed an analytical approach to provide relatively fast estimates of the structure function and its associated comprehensive budget of errors, including the statistical, systematics and sample variance for an ideal case of uniform and isotropic turbulence. \\
\indent With this third instalment in the series, we aim to propagate this formalism down to the reconstruction of the turbulent power spectrum to quantitatively assess the ability to constrain its properties from X-ray measurement. In the present work, we focus on the measurement of centroid shifts of X-ray emission line from the ICM. We apply our method to the future \textit{Athena}/X-IFU instrument in order to quantify the feasibility of this key scientific objective to the mission as a function of the instrument performances. \\
%
\indent The paper is organized as follow: In Sect.~\ref{s:form}, we introduce the basic formalism on the line shape and structure function. We present our method to mock X-ray observations as well as their post processing analysis in Sect.~\ref{s:mock}. In Sect.~\ref{s:err}, we provide the details of the structure function model and complete error budget computation. We discuss the results of our end-to-end simulations and analysis down to quantitative constraints on the turbulent power spectrum in Sects.~\ref{s:res} and \ref{s:disc}. We present our conclusions in Sect.~\ref{s:conc}. \\
\indent The work presented here has been done with the current X-IFU instrumental configuration. However, a reformulation of the Athena mission has been led in an effort to meet budgetary requirements. A new instrumental configuration is yet to be confirmed and future work will need to be done to assess the impact of this new configuration on the work presented here. We assume a $\Lambda$CDM cosmology, with $H_0=70$\,km\,s$^{-1}$\,Mpc$^{-1}$, $\Omega_m=0.3$ and $\Omega_\Lambda=0.7$. With this setup, at a redshift of $z=0.1$, 1 kilo-parsec (kpc) corresponds to an angular extend of 0.54~arcseconds (arcsec).

\section{Structure function and its uncertainty} \label{s:form}

The two most common direct diagnostics available to study turbulence from X-ray spectroscopic observations, via gas motions along the line of sight (LoS), are the line centroid shift from its rest frame value and the line broadening \citep{Cucchetti2019}. In this work we focus on the centroid shift measurements over a given observed area. We assume that the statistics of this observable and, thus, of the underlying velocity, can be characterized using the second-order structure function, which is closely related to the power spectrum. In other words, we assume the turbulent field to be closely approximated by a Gaussian random field. 

\subsection{Second-order structure function} 

To characterize the gas turbulence as a function of different scales, we use the second order structure function SF(\textbf{r}). It is equivalent to a spatial two points correlation function and appears when observing the distribution of the velocity field over all the points in space. We assume that the velocity field, \emph{v,} is isotropic and for any given component of the velocity field, we define a structure function such that:
\begin{equation}
    SF(\mathbf{r}) = \overline{( \emph{v}(\mathbf{x} + \mathbf{r}) - \emph{v}(\mathbf{x}) )^2 },
\end{equation}
where the over-line denotes averaging over the different pairs of sky points \textbf{x} and \textbf{x}+\textbf{r}, separated by a distance, $r = |\mathbf{r}|$. \\
\indent In practice, only the \AM{LoS} component of the velocity field is accessible from spectroscopy measurements, modulated by emissivity effects. We can, however, transpose the structure function definition to its 2D projected equivalent by using the centroid shift as a velocity proxy. \\
\indent As developed in \cite{Clerc2019}, in the case where line diagnostics are obtained from spectra measured over 2D bins (simply due to the pixel size or because of spatial binning over larger bins), the structure function can be written as :
\begin{equation} \label{eq:sf_grid}
    SF(s) = \frac{1}{N_p(s)} \sum_{\sepa} |C_{\wwp} - C_{\ww}|^2
,\end{equation}
where $\sepa$ is the distance separating any pair of bins, $\ww$ and $\wwp$, while $N_p$ is the number of pairs separated by the distance, $\sepa$, and $C_{\ww}$ is the observed centroid shift in velocity space measured over a given emission line or spectrum,  obtained from the flux-weighted sum of all the single lines of sight contributing to a given region (pixels or bins). \\
\indent The measurement of the structure function provides a view of the underlying turbulent velocity power spectrum, and can be used to estimate some characteristic parameters of the turbulence such as injection and dissipation scales, the slope of the turbulent cascade, or its normalization factor. \\
\indent A key point however, is that the realization of a turbulent flow is naturally stochastic. Thus, a given turbulence regime, characterized by its power spectrum, can produce different velocity fields and thus different measured structure functions if the observed field of view is limited. This principle is called sample (or cosmic) variance. The structure function $SF$ is a random variable, which depends on the particular realization of the velocity field. We are interested in its mean value and variance across multiple realizations, which are  denoted as$\meansf$ and $\sigma_{sf}^2$, respectively. \\
\indent In addition to sample variance, the limited exposure time of observation and uncertainties in the fit or in the calibration lead to statistical and systematic errors in X-ray observations of the centroid shifts.
We use $\sigma_{stat}$ to denote the standard deviation of the statistical error for $C_{\ww}$. 
This uncertainty adds scatter to the velocity measurements on top of the sample variance. \cite{ZuHone2016} showed that any measurement of the structure function is systematically biased by the statistical error, provided that statistical errors are not correlated with the value of $C_{\ww}$, such that:
\begin{equation} 
    \label{sig_stat}
    \meansf^{\prime} = \meansf + 2 \sigma_{stat}^2.
\end{equation}
One way to understand this result is that the structure function provides a measurement of the power of fluctuations at different scales, in the same way the power spectrum does. In that sense, a uniform measurement error can be seen as another source of fluctuations and adds bias to the SF as such. This formula holds, as long as all bins have independent and identically distributed measurement uncertainties.

\subsection{Estimation of the full error budget} 

\cite{Cucchetti2019} showed that the variance of the structure function is also systematically biased by terms related to the statistical error on the centroid shifts measurements. They obtained:
\begin{equation} \label{full_error}
    \sigma_{sf, {\rm tot}}^2 = \sigma_{sf}^2 + \sigma_{vf}^2 + \sigma_{sff}^2 + \sigma_{statf}^2
,\end{equation}
\AM{Let us detail the terms in the equation above. $\sigma_{sf}^2 $ the intrinsic variance of the structure function. This is the residual variance in the absence of statistical error, namely, if $\sigma_{stat}$=0. \\
The velocity field fluctuation term, $\sigma_{vf}^2 = 4\, \sigma_{D}^2\, \sigma_{stat}^2$,  is related to the velocity fluctuations in the field of view. The larger the pixel-to-pixel variations in the velocities, the larger this term. This term is small for dissipation scales larger than the pixel or binned region size. Also, $\sigma_{D}^2$ is the variance of the first order structure function (as defined in Eq.~\ref{eq:d_definition}) \\
The structure function fluctuation term, written such that  \\ $\sigma_{sff}^2 = \frac{4}{N_p(s)}\, SF(s)\, \sigma_{stat}^2$, is related to the uncertainty with which the structure function is computed when the turbulent velocities are affected by statistical errors. This term is negligible if the number of pairs, $N_p({\sepa}),$ used to estimate the structure function is large. \\
The statistical fluctuation term, which is written such that $\sigma_{statf}^2 = \frac{4}{N_p(\sepa)}\, (N_{nei}(\sepa) + 1)\, \sigma_{stat}^4$ , is purely a contribution of the statistics to the overall variance of the structure function.  
This term is also negligible if the number of pairs used to estimate the structure function is large, but otherwise it is most important at small spatial separations. Then,
$N_{nei}(\sepa)$ is the number of neighboring regions at a distance $\sepa$ of any given point.}

\section{Mock X-ray observations} \label{s:mock}

One of the key science objectives of Athena/X-IFU is to understand the role of turbulence in the gas dynamical assembly and relaxation in massive halos, namely, groups and clusters of galaxies. In order to assess this capability, we perform end-to-end simulations using the SIXTE software \citep{Dauser2019}. The following sections describe our procedure, as a continuation of the work of \citet{Cucchetti2019}, with the difference that the present simulations are based on a physically motivated input cluster model to Athena/X-IFU mock observations.

\subsection{Galaxy cluster model}

\subsubsection{General characteristics}

\indent Nearby clusters are targets of choice in order to grasp the whole range of spatial scales spanned by turbulent cascades. We modeled a nearby galaxy cluster with a cosmological redshift $z=0.1$ and a scale radius of $R_{500}=1300$\,kpc, corresponding to an angular radius of $11.7$\,arcmin (where $R_{500}$ is the radius encompassing a mean density of the gas 500 times that of the critical density of the Universe at the cluster redshift). 

\indent We assumed a non-cool core cluster model. Systems in this class are more likely to have undergone major events such as cluster mergers or collision and are more likely to show a higher level of bulk motions and turbulence. 
Our cluster gas model departs from the popular, albeit simplistic, $\beta$-model. We remain under the assumption of a spherically symmetric body.
We only include instrumental background as an additional component in our simulations, neglecting astrophysical back- and foreground emission as well as any second-order instrumental systematic.

\subsubsection{Profile}

The cluster density and temperature profiles are taken from the joint analysis of eight non-cool core clusters in the XMM-{\it Newton} Cluster Outskirts Project \cite[X-COP][]{Ghirardini2019}.
This temperature profile follows the parametrisation given by:
\begin{equation}
    \dfrac{T(x)}{T_{500}} = T_0 \dfrac{\sfrac{T_{min}}{T_0} + (\sfrac{x}{r_{cool}})^{a_{cool}}}{1 + (\sfrac{x}{r_{cool}})^{a_{cool}}} \dfrac{1}{\left(1 + (\sfrac{x}{r_t})^2\right)^{\sfrac{c}{2}}}
,\end{equation}
where $x = r/R_{500}$, and \{$T_0, T_{min}, r_{cool} a_{cool}, r_t, c/2$\} are six parameters, and $T_{500}$ follows a parametrization on the redshift, see \cite{Ghirardini2019}. Their values are provided in Table \ref{XCOPparameters}. 

\begin{table}
    \centering
    \begin{tabular}{|c|c|}
       \hline
        Parameter & Value\\
        \hline \hline
        $T_0$ & 1.09 \\
        $T_{min}/T_0$ & $0.66$\\
        $\log(r_{cool}) $ &  -4.4 \\
        $a_{cool}$ & 1.33 \\
        $r_t$ & 0.45  \\
        $c/2$ &  0.30 \\
        $\gamma$ & 3\\
        $\log(n_0)$ [$\text{cm}^{-3}$] & -4.9 \\
        $\log(r_c)$ [kpc] & -2.7 \\
        $\log(r_s)$ [kpc] & -0.51 \\
        $\alpha$ &  0.70 \\
        $\beta$ & 0.39 \\
        $\epsilon$ & 2.60 \\
         \hline
    \end{tabular}
    \caption{Values of the parameters used for the temperature and emission models, obtained from the best-fit of these profiles to observed clusters \citep{Ghirardini2019}.}
    \label{XCOPparameters}
\end{table}
The density profile follows: 
\begin{equation}
    n_e^2(x)= n_0^2 \dfrac{(\sfrac{x}{r_c})^{-\alpha}}{(1 + (\sfrac{x}{r_c})^2)^{3\beta -\alpha /2}} \dfrac{1}{(1 + (\sfrac{x}{r_s})^{\gamma})^{\sfrac{\epsilon}{\gamma}}}
,\end{equation}
where x = r/$R_{500}$ and \{$\gamma,n_0, r_c, r_s,\alpha,\beta,\epsilon$\} are six parameters. Their values are provided in Table \ref{XCOPparameters}. We also set an arbitrary upper limit on the electron density value at 0.05\,cm$^{-3}$, in order to avoid divergence as we go closer to the center of the cluster.

For the  metallicity profile, we followed the shape derived by \citet{Mernier2017}, which was obtained from the study of a sample of 44 nearby cool-core clusters, groups, and ellipticals observed with XMM-{\it Newton}:
\begin{equation}
    Fe(r) = 0.21 (r + 0.021)^{-0.48} - 6.54\times \exp\left(- \frac{(r + 0.0816)^2}{0.0027}\right)
,\end{equation}
where $r$ is the distance from the center of the cluster in kiloparsecs. 
The solar abundances are fixed to those of \cite{1989GeCoA..53..197A}.

We assume an emission spectrum for the intra-cluster gas described by  the APEC model \citep{2001ApJ...556L..91S} implemented under XSPEC \citep{1996ASPC..101...17A}. It assumes an ionized, collisional, diffuse, and optically thin plasma.  
As we assume a fully ionized gas with $\sim$10\% Helium in atom number (and weakly contributing heavier elements),  electron density, $n_e$ and hydrogen density, $n_H$, are related by $n_e \sim 1.2 \: n_H$. We imposed the emissivity of the gas to vanish at radii beyond $5 R_{500}$.

\subsubsection{Turbulent velocity}

We assumed that the turbulent field is homogeneous and isotropic. Following \cite{ZuHone2016}, we modeled the 3D velocity power spectrum with a standard functional, with $k_{\text{inj}}$ and $k_{\text{diss}}$ as the cut-off frequencies of the Kolmogorov cascade at low and high wave numbers, $\kthree$ (with $k=1/r$ the conversion between scales and wave numbers), $\alpha$  as the slope of the spectrum, and $\sigma$ as the 3D velocity dispersion (units km\,s$^{-1}$). With $k = |\kthree|$, our model can be expressed as:
\begin{equation}
\label{P3Deqn}
    P_{3D}(\kthree) = \sigma^2 \frac{\displaystyle k^{-\alpha} e^{-(k_{\text{inj}}/k)^2} e^{-(k/k_{\text{diss}})^2}}{\displaystyle \int 4 \pi  k^2 k^{-\alpha} e^{-(k_{\text{inj}}/k)^2} e^{-(k/k_{\text{diss}})^2} dk}
.\end{equation}
\AM{In our simulations, the injection scale is set to 300\,kpc. While it has not been determined by observations, we expect similar values if turbulence is driven by mergers or accretion from the large-scale structure \citep[as found in][]{Dupourque2022}. The dissipation scale is set to 10\,kpc. The exact range of it can depend on the details of the ICM dynamics and properties and can vary between 1 to 100\,kpc \cite{2011ApJ...742...19C}. It was chosen to be 10\,kpc arbitrarily with the perspective of testing X-IFU capabilities in this regime. The slope is $\alpha = -11/3$, corresponding to the usual index of the Kolmogorov cascade in the energy spectrum. The normalization constant, $\sigma,$ is set such that the expected velocity broadening of the turbulent field at the center of the cluster equals $(\mathcal{M} c_s)^2$. Here, $\mathcal{M}$ is the Mach number, set to 0.3 following \cite{ZuHone2016}, and $c_s$ is the speed of sound, set to $833$\,km/s. This gives $\sigma \simeq 250$\,km\,s$^{-1}$. }

We generated a realization of the line of-sight turbulent velocity field to match such a power spectrum by drawing a random field in the Fourier domain, $V_x(\mathbf{k}) = v e^{i\phi}$. To do so, we used a Rayleigh distribution for the amplitude, $v$, and a uniform distribution for the phase, $\phi$. 
The inverse Fourier transform of $V_x(\mathbf{k})$ gives the velocity field \citep{Cucchetti2019}. 
The turbulent motions of the gas are then included in our cluster model by converting the \AM{LoS} component of the velocity field into a redshift correction applied to the emitted spectra.

\subsection{Post-processing} 

\subsubsection{Mosaics}

To fully map our toy model cluster at z = 0.1 out to $R_{500}$, a coverage of 19 separate pointings arranged around the center of the cluster is needed. We set the initial exposure time of each pointing to 500 ks. Each pointing mock observation is simulated separately. This relatively high exposure time is not meant to be performed on 19 different exposures during the mission.  From this main simulation, we are able to pick the number of pointings and their individual exposure time by downsampling the observations to the desired length and by assembling the final event list suited to a given observing strategy.
From the list of events from the SIXTE simulation, we reconstructed raw brightness maps in counts (later referred to as count maps).

\subsubsection{Spatial binning} 

In order to reconstruct the line-of-sight-projected velocity field, we binned the 2D counts distribution of our mosaic mock observations into bins of multiple pixels. The binning was performed over multiple pointings of the same exposure times. We aim to define regions of equal signal-to-noise ratios (S/Ns) to extract individual centroid shift measurements  with similar levels of statistical quality.

In practice, this binning procedure operates on the assembly of count maps with similar exposure time, dominated by the ICM Bremsstrahlung emission, and the bins (denoted as $\mathcal{W}$)  are created following a Voronoi tessellation \citep{2009arXiv0912.1303C} with a constant S/N of 200.

\subsubsection{Spectral fitting} 

For each of the Voronoi bins, we extracted a spectrum generated from the list of events. 
Each spectra is fitted under XSPEC, using the APEC model. More precisely, the spectra were fitted to an absorbed single-temperature thermal plasma model, including a velocity broadening component of the lines to account for turbulent velocities and accounting for the absorption along the \AM{LoS} (i.e., \texttt{phabs*bapec}). The rebinning of the spectra was done before the fitting to increase the computation speed, following the optimal binning proposed by \citet{2016A&A...587A.151K}.

The column density, $n_{abs}$, of the \texttt{phabs} model is kept fixed to the value of $0.03 \cdot 10^{22} \text{cm}^{-2}$. The free parameters are the temperature, the abundance, the normalization, the redshift, and the line velocity broadening.
All the parameters are fit simultaneously, over the full energy band (0.2 – 12 keV), and relying on C-statistics for the goodness of fit.

In order to account for the vignetting effects in our processing, a specific ARF is created for each bin, following the method by \cite{Cucchetti2018}. This is done by multiplying the response of each pixel within the bin by the vignetting function implemented in SIXTE, over all energies, and averaging, in each energy channel, the response of each pixel weighted by their respective number of counts.

\subsubsection{Output and input parameter maps} 

After fitting, we used the best fit values and errors inside each bin to create output maps for each parameter and its associated error. Each spatial bin is associated to a value of the centroid shift, $C$, thus creating a spatial map of this parameter. The centroid shift, $C,$ (tracing the turbulent motions), however, needs to be converted from the best fit value of the spectral redshift, $z_{fit}$, to velocity domain with :
\begin{align}
    z_v &= \frac{z_{fit} - z}{1 + z} ,\\
    C &= \frac{(1 + z_v)^2 - 1}{(1 + z_v)^2 + 1} \label{z_to_c}
 \: c_{light},
\end{align}
where $z_v$ is the redshift equivalent to the integrated \AM{LoS} velocity induced by the injected turbulent motion in our simulations, $z$ the cosmological redshift of the cluster, and $c_{light}$ the speed of light. Equation \ref{z_to_c} can be approximated by $C = z_v c_{light}$ as the turbulent velocities are non relativistic. Our procedure relies on the assumption that the cosmological redshift of the cluster is perfectly known.

A map of the true centroid shift is created and serves as a reference to assess the goodness of fit of our spectral analysis. We obtain this map by projecting the velocity shift along the \AM{LoS}, weighted by the emissivity of the ICM \citep[see Sect.5.4 in][]{Cucchetti2018}.

\subsubsection{Measured structure function}

The second-order structure function was computed from the output map of measured centroid shifts, $C$, following Eq.~\ref{eq:sf_grid}. The associated statistical errors arising from the propagation of the best-fit parameter uncertainties were evaluated with the difference between the input and output bulk motion maps. This statistical error is then saved for subsequent application in the computation of the structure function variance.

\section{Model and fit of the structure function} \label{s:err}

In order to model the measured second order SF, we applied and extended the formalism developed by \cite{Clerc2019} and \cite{Cucchetti2019} to our toy model 
This formalism provides a semi-analytical way to estimate the full error budget over the structure function without having to reproduce the full simulations process a large number of times to estimate the $SF$ mean and scatter (including the sample variance) at each step of a Monte Carlo fitting procedure.

\subsection{Estimation of the SF}\label{s:sfmodel}

From the formulation of the structure function in Eq.~\ref{eq:sf_grid}, we demonstrate in Appendix~\ref{app:sfgrid} that for any emissivity distribution of the sources, projected over a field of view grouped into bins of any shape and at any position, the expected structure function can be given by:
\begin{multline}
    \label{eq:sf_grid_fourier}
    \langle SF(s) \rangle =  
    \sum_{\kthree} P_{3D}(\kthree) \\
    \times \left( 
    \frac{1}{N_p(s)} \sum_{d(\ww, \wwp) = s} \left| \frac{c_{\epsilon \cdot \ww}(\kthree)}{F_{\ww}} - \frac{c_{\epsilon \cdot \wwp}(\kthree)}{F_{\wwp}} \right|^2 \right),
\end{multline}
\AM{In the equation above, the second sum runs over all pairs, $(\ww, \wwp),$ separated by a distance, $\sepa$, which is fixed by the geometry of the bins. $c_{\epsilon \cdot \ww}$ is the Fourier coefficient of the product, $\epsilon(x,\vec\theta)\, \ww (\vec\theta)$, the emissivity at a location, $x,$ along a \AM{LoS} labeled as $\vec\theta$, multiplied by the window function of the bin, which equals one inside the bin and zero outside. Then, $\vec\theta$ is the 2D vector on the sky plane, designating a unique \AM{LoS}. $F_{\ww}$ is the total flux in the bin denoted as $\ww$.}

In order to speed up the computation, we calculated the term in parenthesis in Eq.~\ref{eq:sf_grid_fourier} only once for the given geometry and binning configuration. It was then saved on the disk and then multiplied by $P_{3D}$ each time a new turbulent model was generated, thereby providing the $\langle SF \rangle$.

For a general emissivity model, it is necessary to compute the 3D Fourier transform of $\epsilon(x,\vec\theta)\, \ww(\vec\theta)$ for each bin $\ww$. In order to accelerate this step, we rely on the following approximation (see details in Appendix~\ref{app:sfgrid}):
we assume that both the flux and the emissivity are independent from the \AM{LoS} chosen within the bin. We can then choose any of those \AM{LoS}s and define it as $\vec\theta_{\rm eff}$, representative of the bin $\ww$ (but changing from bin to bin) such that:
\begin{equation}\label{eq:approx_epsilon}
    \forall\: (\vec\theta\, \in\, \ww)\: :\: \epsilon(x,\vec\theta) \approx \epsilon(x,\vec\theta_{\rm eff}) \equiv \varepsilon_{\vec\theta_{\rm eff}} (x).
\end{equation}

Hence, the simplified and faster computation of $c_{\epsilon \cdot \ww}$ through:
\begin{align}\label{eq:approx_cepsilonw}
    c_{\epsilon \cdot \ww} \approx \text{FT}_{1D}(\varepsilon_{\vec\theta_{\rm eff}} (x))\; \text{FT}_{2D}(\ww(\vec\theta)).
\end{align}
\indent This methodology allows for a fast computation of the structure function for any emissivity model and any observation and geometry. We implemented it to compute the modeled structure function in our MCMC procedure to model the structure function.

\subsection{Estimation of the errors}\label{s:sferrors}

\begin{figure}[h]
    \centering
    \includegraphics[width = 8cm]{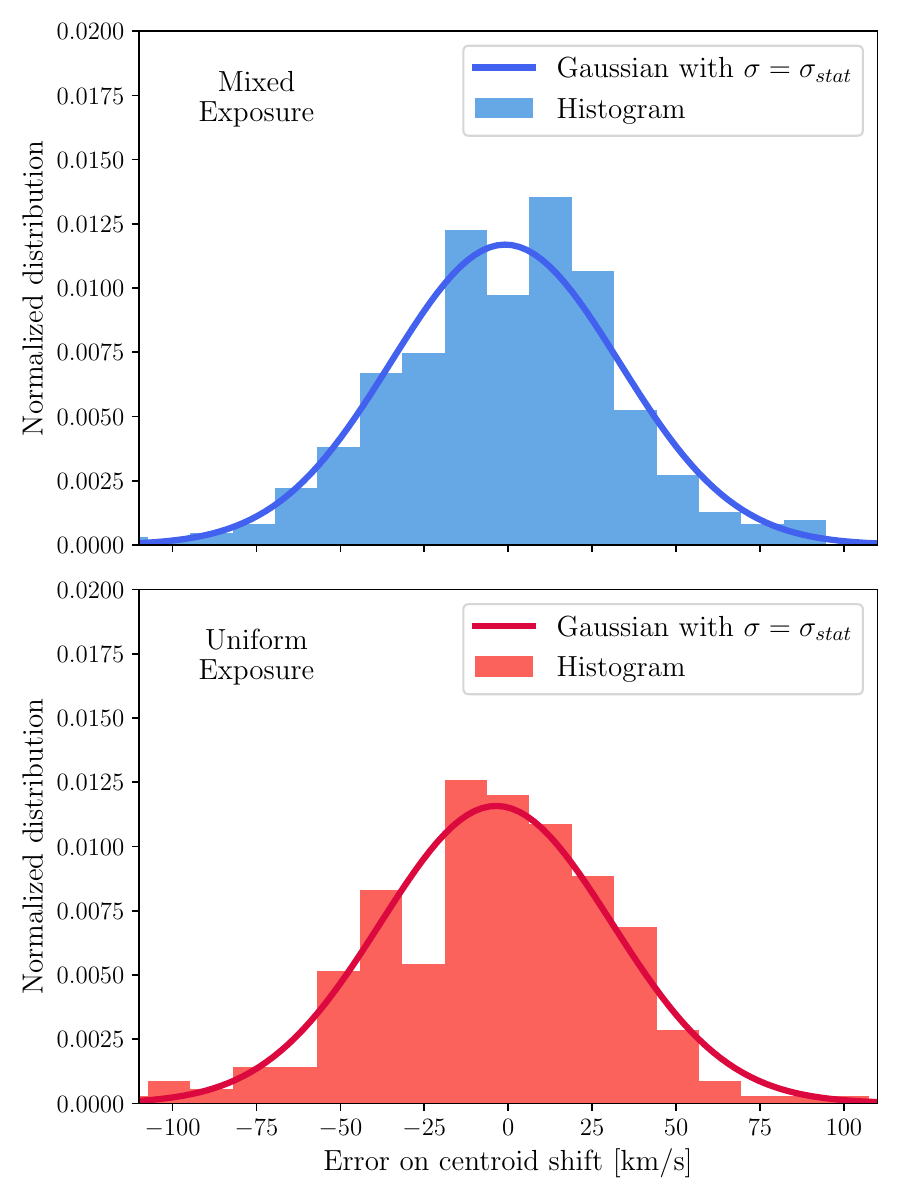}
    \caption{Histograms of the centroid shift measurement error calculated for the mosaic of 19 pointings for both mixed and uniform exposures. The plain lines show the Gaussian profile plotted with a standard deviation corresponding to that of the sample of the errors, that is, 34.1\,km\,s$^{-1}$ for the "mix exposure" and 34.5\,km\,s$^{-1}$  for "uniform exposure."}
    \label{sigmastat}
\end{figure}

The mean structure function, the sample variance, $\sigma_{sf}$, the velocity field fluctuations, $\sigma_D$, can be estimated numerically for any emissivity model, with a few simplifying assumptions on the detector and observation geometry. The other terms needed to assess the full error budget are themselves intrinsically related to the observational setup. These terms are $N_p$, $N_{nei}$, and $\sigma_{stat}$. We note that it is simple to derive $N_p$ and $N_{nei}$, as they are related to the binning and geometry of the detector and can be obtained analytically with high accuracy. \\
\indent We assumed a prior knowledge of $\sigma_{stat}$, which we obtained through a comparison of the output maps with the input maps. For a real observation, this quantity would be hard to estimate and would probably require  a complete end-to-end simulation to get a proper estimate. Figure~\ref{sigmastat} shows the distribution of the measurement errors on the centroid shift, for two observation configurations: the "uniform exposure" and "mix exposure," which are explicited later. The distributions are close to Gaussian, which justifies the correction explicited in Eq.~\ref{sig_stat}.

\subsubsection{Sample variance $\sigma_{sf}$}

Computing the analytic sample variance involves integrals over the 3D Fourier space that make the full computation numerically untractable, at least in the context of multiple evaluations as in an MCMC procedure. However, \cite{Clerc2019} shows a simplifying assumption that allows for a fast computation of the sample variance term. It requires the assumption of  constant emissivity over the sky, as in~\citet{2004A&A...426..387S}, for instance, with a so-called Xbeta model for the emissivity (defined in Appendix~\ref{app:xbeta}). 
This allows for a decorrelation of the emissivity field and the Fourier transform of the emissivity-weighted centroid shift along the \AM{LoS}. Under this approximation, and assuming a regular grid of fixed size bins, the variance of the structure function is expressed as:
\begin{equation}
\label{VarSFeqn}
    Var[SF(\sepa)] = 16 \pi \int_0^\infty \frac{1}{\fovarea} P_l^2(\xi) P_{2D}^{\infty 2}(\xi) (1 - J_0(2 \pi \xi \sepa))^2 \xi \dd \xi 
\end{equation}

\noindent with $P_l$ the power spectrum of the bins of characteristic size, $l$ (square of the Fourier transform of the window function), $P_{2D}^\infty$ as the 2D power spectrum of the centroid shift, extrapolated for an infinite analysis region, $\fovarea$ as the area of the observed region, $J_0$ as the 0-th order Bessel function. One expression for $P_{2D}^\infty$ is :
\begin{equation}
    P_{2D}^\infty = 2 \int_0^\infty P_{3D}(\sqrt{k_x^2 + \xi^2})(1 + 2 \pi k_x R_c)^2 e^{-4 \pi k_x R_c} \dd k_x
,\end{equation}
with $ P_{3D}$ as described in Eq.~\ref{P3Deqn} and $R_c$ the core radius of the Xbeta model. The integrals are computed using the double exponential quadrature, allowing for a fast computation of each integral with a few hundreds of integrand evaluations \citep{takahasi1974double}.

Because the observed velocity field is not collected over a regular grid, but in spatial bins of varying sizes and shapes, we made another assumption. We assumed that to each separation, $\sepa,$ we can associate an equivalent bin geometry and Xbeta core radius to be the input of Eq.~\ref{VarSFeqn}. This equivalent geometry is obtained by identifying the bins involved in the computation of the structure function at the separation defined by $\sepa$ and assuming that they can be approximated by a grid of identical pixels of fixed size. Their size is then computed with the mean of the sizes of the pixel used in the separation, $\sepa$. The Xbeta core radius is found by fitting the structure function provided for the Xbeta model in \cite{Clerc2019}, to the structure function computed for the full toy model, as described in Sect. \ref{s:sfmodel} at the true parameters. In that sense, we adapted the Xbeta model to provide the best possible estimation of structure function (and, hence, the variance) around the true value of the turbulent parameters. Finally, the area, $\fovarea$, associated to each separation, $\sepa,$ is taken as the area covered by all bins entering the calculation of $SF$ at this separation. This approach assumes that there is no covariance between the different scales of the structure function. Implementing this effect in our current framework would require us to estimate the covariance over the structure function points, each being estimated on a different grid of separations. This issue is beyond the scope of this study.

To gain insights into the actual dependence of sample variance errors on the various model parameters, we provide in Appendix~\ref{app:sf_scalingerr} an approximate derivation of the fractional uncertainty $\sigma_{sf}/\meansf$. This shows in particular that relative uncertainties are inversely proportional to the largest scale spanned by the observation and roughly proportional to the injection scale in the regime of small spatial bin sizes. The binning size is also relevant, as larger bins will wash out the signal and tend to increase the relative uncertainties. Exact derivations are intractable analytically and would require a numerical integration.

\subsubsection{Velocity field fluctuation term, $\sigma_D$}

The computation of this term calls for simplifying assumptions to be adopted as well. The detailed computation is explicited in Appendix~\ref{app:vard}. We can summarize it in the following way: $\sigma_D$ is computed once for a given observation geometry (i.e., binning) for the true turbulent parameters. It is then saved and later approximated as a term proportional to the structure function when called in the computation of the errors in the MCMC. This is a crude approximation reflecting the similarity between the expressions for $\meansf$ and $\sigma_D^2$, both depending linearly on the 3D power spectrum. This can be explicitly expressed in the following way:
\begin{equation}
    \left.\sigma_D^2(\sepa)\right|_X = \left.\sigma_D^2(\sepa)\right|_{X_0} \frac{\left.\langle SF(\sepa) \rangle \right|_X}{\left.\langle SF (\sepa)\rangle \right|_{X_0}}
,\end{equation}
where we call $X$ the set of turbulent parameters and $X_0$ the set of the true turbulent parameters.

\section{Results}
\label{s:res}

\subsection{Outputs from cluster simulation} \label{s:2.3.2.1}

\begin{figure}
    \centering
    \includegraphics[width = 8cm]{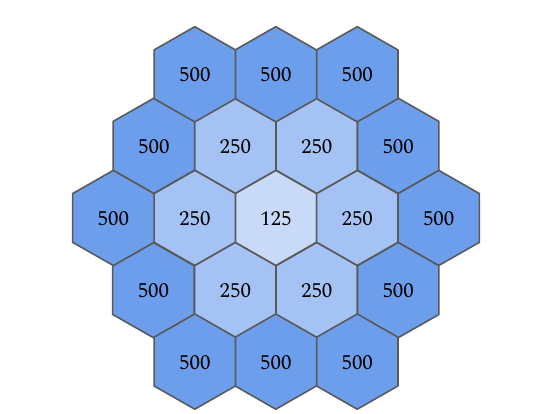}
    \caption{Illustration of the 'mix exposure' pointing strategy with exposure times in ks at the center of each pointing.}
    \label{fig:2.3.2.1.out_counts}
\end{figure}

\begin{figure}
    \centering
    \includegraphics[width = 9cm]{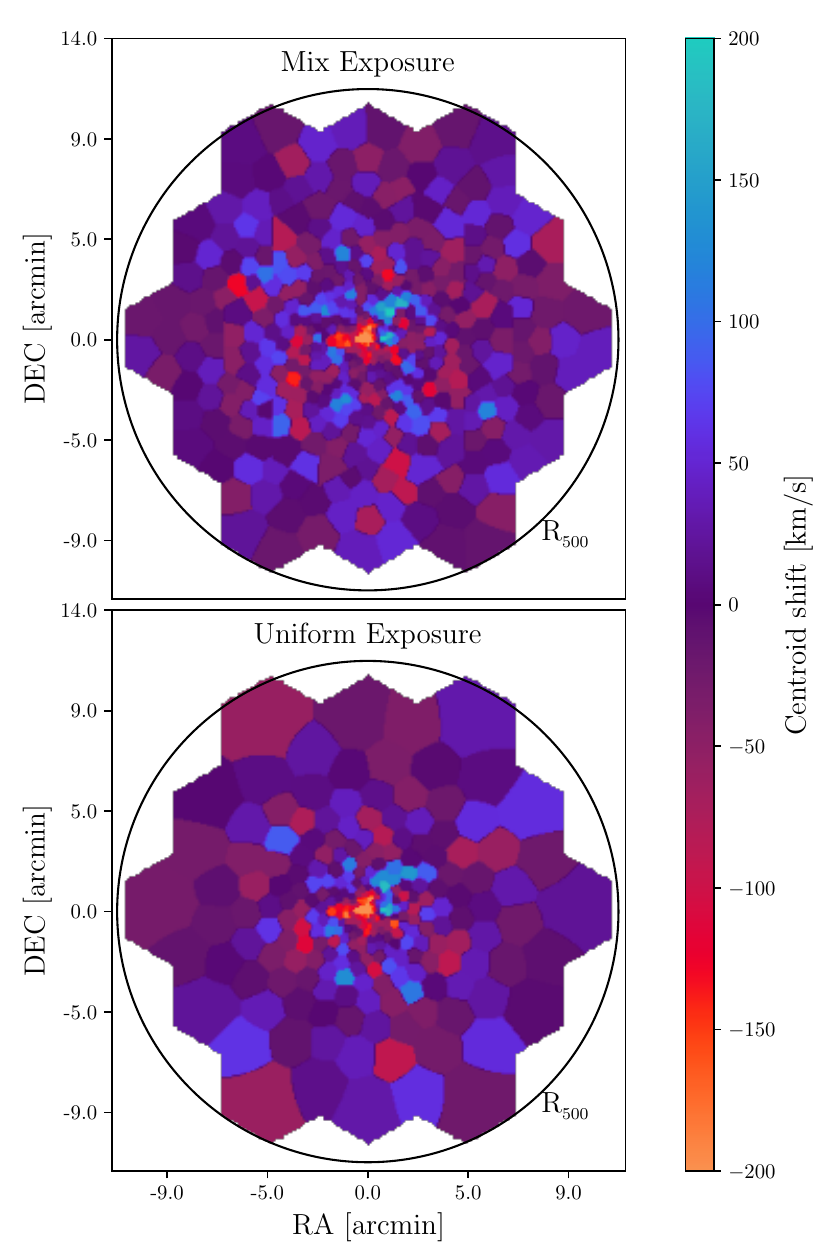}
    \caption{Output maps for mosaics with a mix exposure (top) versus constant exposure at 125 ks per pointing (bottom), for the centroid shifts. A black circle indicates the $R_{500}$ radius. All the spatial bins were constructed such as to provide identical signal to noise levels in the resulting spectra. }
    \label{fig:2.3.2.1.out_centroids}
\end{figure}

\indent For the simulations of 19 pointings, two different mosaics have been produced: one where all the pointings have the same exposure time of 125~ks (for a total exposure of 2.3~Ms) and one where the exposure time was increased toward the outskirts of the cluster by having the central pointing at 125 ks, the first ring at 250 ks, and the second ring at 500 ks (meaning a total exposure of 7.6 Ms, see Fig.~\ref{fig:2.3.2.1.out_counts}). The reasoning for the mix exposure version is an attempt to mitigate the decrease in count rate and, hence, an increase in bin sizes, as the emissivity decreases toward the outskirts. This allows us to obtain output maps with more homogeneous bin sizes (see Fig.~\ref{fig:2.3.2.1.out_centroids} for the centroid shift). \\
\indent Such exposure times are obviously prohibitive and not realistic, so they would not be used for actual observations with the X-IFU. They have been used in order to provide a first configuration with low statistical errors to validate this simulation framework. Further tests, looking at optimizing the observing strategy for future X-IFU observations and, thus, with more realistic total exposure times, are planned in the near future. Furthermore, the hypothesis of isotropy of turbulence would allow for the use of only a quarter or a slice of our 19 pointings, offering a significant time reduction. 

From the output and input maps of the centroid shifts, we generated the histogram of the absolute errors from individual regions (see Fig.~\ref{sigmastat}). The standard deviation correspond to the $\sigma_{stat}$ from Eq.~\ref{sig_stat}, used for the computation of the expected structure function and error budget. For both exposure configurations, we find a statistical deviation of  $\sim$30\,km\,s$^{-1}$.

\subsection{Simulated SF versus modeled}

From the output maps of the measured centroid shifts, we extracted the corresponding structure functions. From these,  we subtracted the appropriate bias correction dependent on $\sigma_{stat}$. 

\begin{figure}[h]
    \centering 
    \includegraphics[width = 9cm]{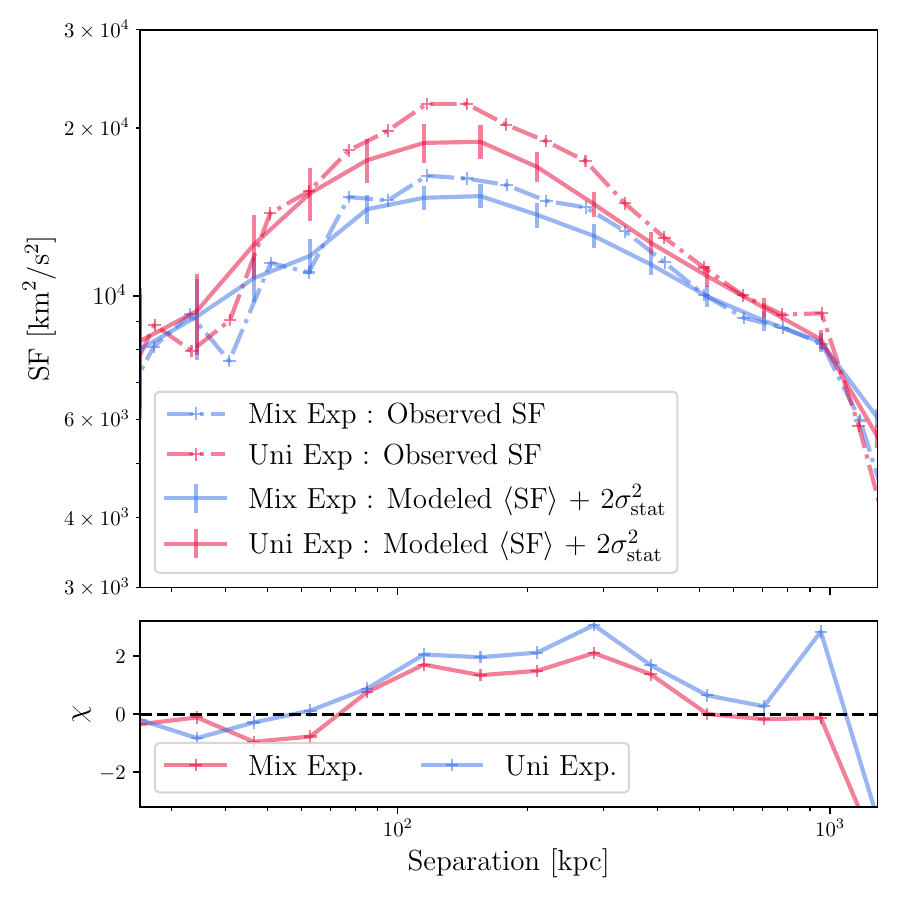}
    \caption{Structure functions and residuals for our simulation of 19 X-IFU pointings. The blue curves show the results for the "uniform exposure" configuration and the red curves refer to those of the "mix exposure" configuration. The upper plot shows the structure functions, observed in the mock observations in dash-dotted lines, and the modeled in plain lines. The total error is over-plotted on the modeled structure function with error bars, as its computation is part of the modeling, as described in Sect. \ref{s:sferrors}. The lower panel shows the value of  $\chi =$(data-model) and the errors.}
    \label{ObservedSFs}
\end{figure}

\begin{figure}[h]
    \centering 
    \includegraphics[width = 9cm]{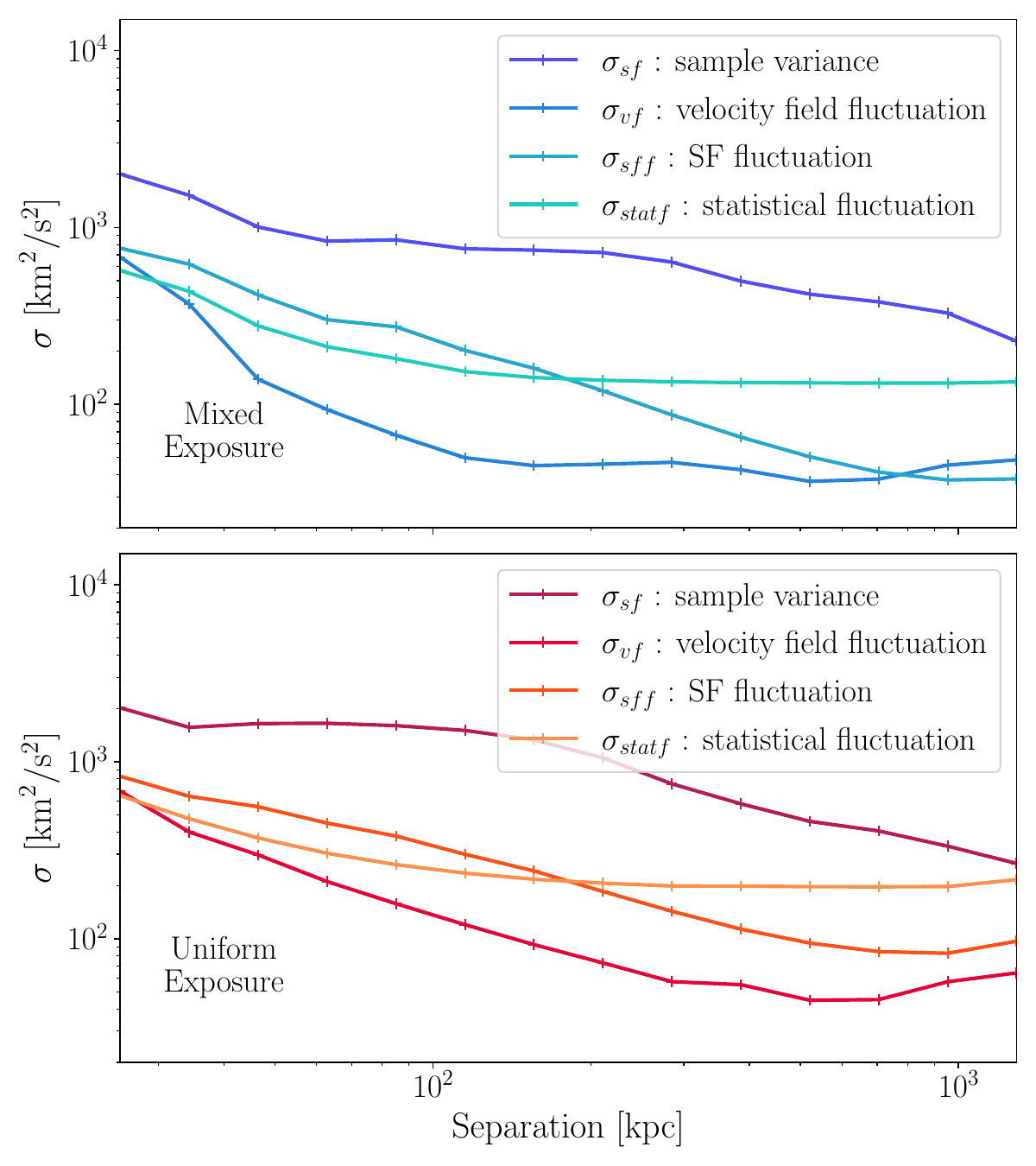}
    \caption{Plot of the error terms contributing to the total error computation as a function of separation. The top panel represents the error models for the "mix exposure" computed with the input turbulent parameters. The lower panel represents these same errors computed for the "uniform exposure." Four terms are represented, going from darker  to lighter shades: the sample variance, the velocity field fluctuation, the structure function fluctuation, and the statistical fluctuation.}
    \label{VarSFcontributions}
\end{figure}

\indent Figure \ref{ObservedSFs} shows a comparison of structure functions from the mock observations (dashdotted lines) and from modeling (plain lines with error bars), for the uniform exposure configuration (in red) and mixed exposure (in blue). We stress that computation of the uncertainties are entirely part of the model. This explains why we prefer representing error bars on top of the model SF, rather than the measured SF. The overall shape and normalization of the modeled structure functions match with those from the mock observations. The central part of the SF, between 70 and 300\,kpc separations, shows that the model underestimates this specific realization of the structure function. It is partly due to this specific realization of the turbulent velocity field, as from its stochastic nature another realization would show a different deviation; and partly to the hypotheses underlying the computation of the theoretical structure function. In the uniform exposure, where the bins are larger, the approximation made in Equations~\ref{eq:approx_epsilon}--\ref{eq:approx_cepsilonw} is more likely to fail and to result in a stronger mismatch with respect to the data. \\
\indent We illustrate the different contributions of the error terms to the total error in Figure~\ref{VarSFcontributions}. The sample variance dominates at all scales. The other terms have comparable contributions at lower scales. At higher scales, the statistical fluctuations become comparable with the sample variance. All of the error terms but the cosmic variance include a contribution of the measurement error $\sigma_{stat}$. This means that higher measurement errors would lead to these other terms having a larger contribution compared to the sample variance.   We recall that, in principle, $\sigma_{stat} \propto \frac{1}{\mathrm{SNR}} \propto \frac{1}{\sqrt{t_{exp}}}$.  This is true if we keep the bin size constant. In our specific case however, the high spectral resolution of X-IFU requires a rather high (i.e., approx. 200) S/N to populate the numerous X-IFU spectral channel in order to characterize emissions lines. This forces us to keep the SNR constant and to adapt the binning accordingly.  Then, even with lower exposure times, the $\sigma_{stat}$ term should remain constant. As a result, the relative contributions of the error should remain relatively unchanged. The resulting structure function would span a different range of distances because of the change in bin size to accommodate for the lower exposure time. The overall error level should increase, as all the terms are dependent on the bin size, but their relative contributions should remain on the same order.

\begin{figure}[h]
    \centering 
    \includegraphics[width = 9cm]{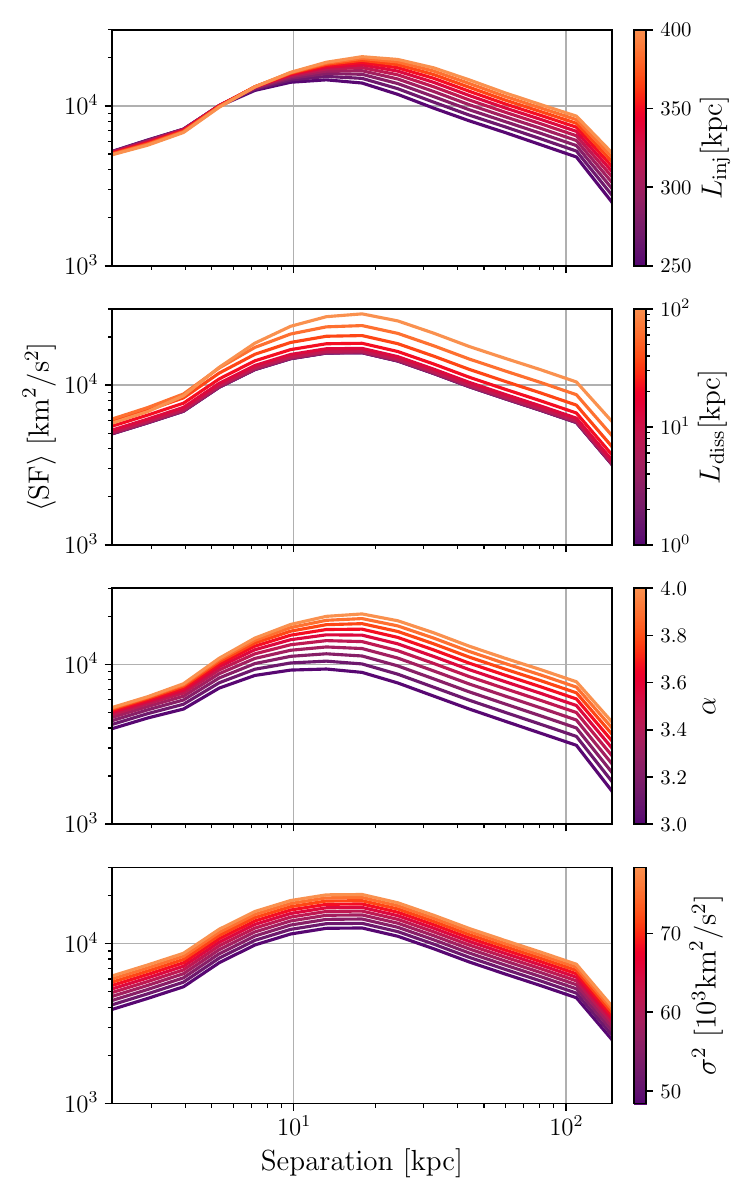}
    \caption{Structure function models for our simulation of 19 X-IFU pointings. The red curves show results for the "uniform exposure" configuration}
    \label{VaryingSFs}
\end{figure}

\indent In Figure \ref{VaryingSFs} we show how the modeled structure function varies with the turbulent parameters. The range given to the parameters is roughly representative to the expected range in which they should be observed in the ICM \citep[e.g.,][]{ZuHone2016, Simionescu2019, Vazza2011}. The variations of the parameters have a relatively similar impact on the shape of the structure function. This lead us to expect strong degeneracies in the posterior distributions of the parameters. The dissipation scales ranging between 1 and 10\,kpc have little to no effect on the structure function. This is due to the finite size of the bins, which, at this redshift, are 8.72\,kpc in width. From this we can expect the dissipation scale to be hardly constrained in this observational configuration. 

\subsection{Retrieving the turbulent parameter distribution}

\indent To obtain the constraints on the turbulent parameters of the ICM, we used the \texttt{emcee} python package to perform a Monte Carlo sampling of the posterior distribution of the parameters. We ran 32 walkers for 10000 steps each. We checked proper convergence of the chains with the $\hat{R}$ statistic \citep{2019arXiv190308008V}, which is 1.01<$\hat{R}$<1.06 for all parameters. This is not the optimal value but was deemed acceptable with respect to the computation time needed for a greater number of samples. We used a Gaussian likelihood, with errors depending on the turbulent parameters, resulting in the following expression: 
\begin{align*}
    \log\left(p(SF_{obs} | X)\right) & = -0.5 \sum_{\sepa} \biggl[ \log(2 \pi\sigma_{sf, {\rm tot}}^2) \\ & + \frac{( \langle SF(\sepa, X) \rangle - SF_{obs}(\sepa))^2}{\sigma_{sf, {\rm tot}}^2} \biggr],
\end{align*}
with $X$ as the set of turbulent parameters, $\langle SF(\sepa, X) \rangle$ as the modeled SF described in Sect. \ref{s:sfmodel} and $\sigma_{sf, {\rm tot}}^2$ as the modeled variance of the SF described in Sect. \ref{s:sferrors}. 

\begin{figure}[h]
    \centering 
    \includegraphics[width=9cm]{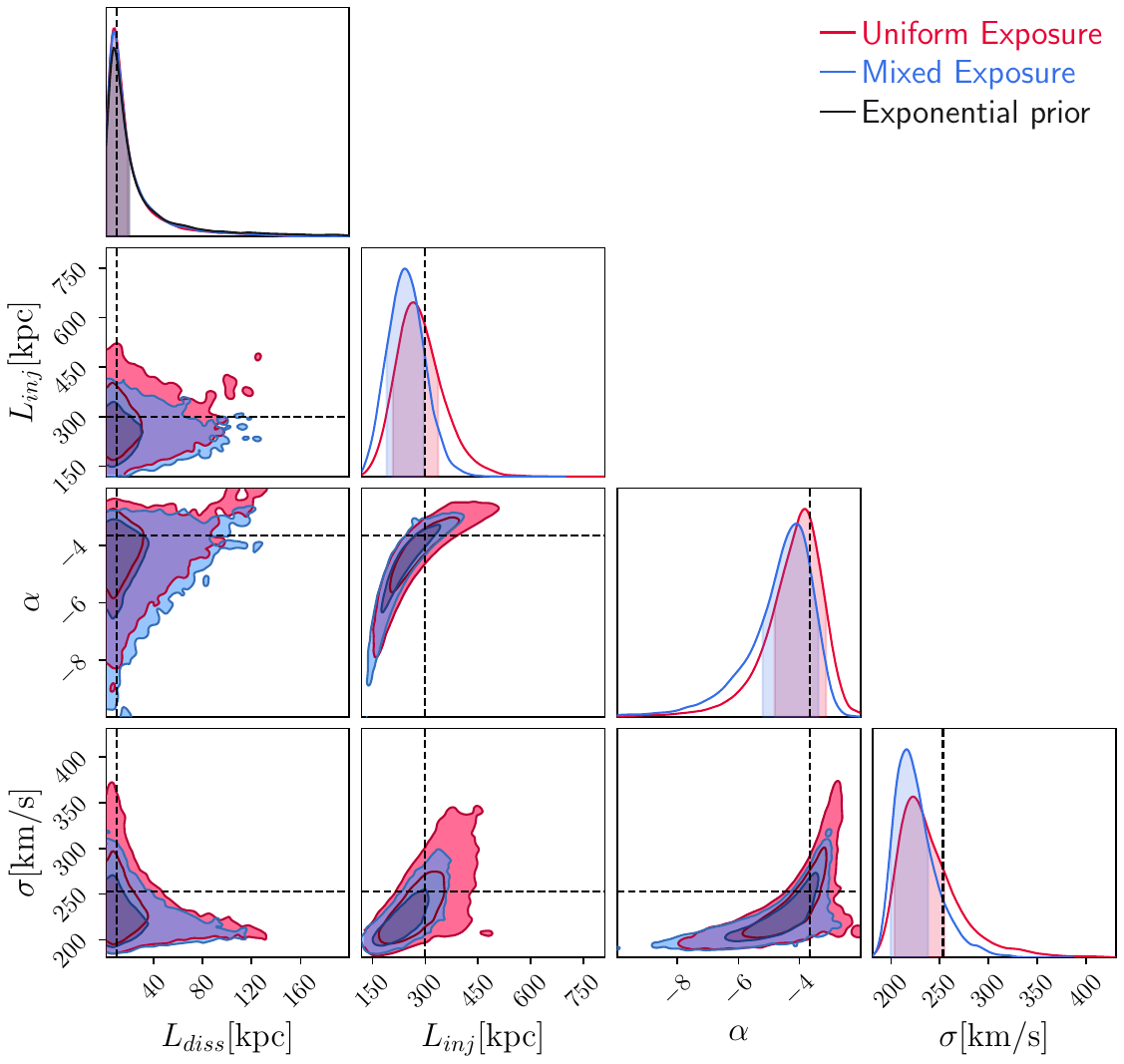}
    \caption{Posterior distribution of the turbulent parameters obtained through the MCMC sampling. The blue contour show the distribution for the mixed exposure observation and the red for the uniform exposure. The inner contour delineates the 1$\sigma$ level of the posterior probability, the second contour is at 2$\sigma$. Dotted lines show the input values of the simulations.  Uniform priors were used for all parameters, except $L_{\mathrm{diss}}$, for which the prior is plotted in black.}
    \label{contour_kdiss_expprior}
\end{figure}

\indent We ran the MCMC sampler for different configurations of priors and free parameters, for both the mix exposure and uniform exposure observations. 
First, we investigated the possibility of constraining all the parameters at once, including the dissipation. We used uniform priors for the following parameters, such that $\pi(k_{\mathrm{inj}}) \sim \mathcal{U}(0,k_{\mathrm{diss}}), \pi(\alpha) \sim \mathcal{U}(-10,0),  \pi(\sigma^2) \sim \mathcal{U}(0,500^2) $. Because $k_{\mathrm{diss}}$ can span the entire range [0;+$\infty$[, we set the prior to the exponential distribution, such that $\pi(k_{\mathrm{diss}}) = \lambda e^{- \lambda k_{\mathrm{diss}}}$. This follows from the exponential distribution being the maximum entropy distribution on the [0;+$\infty$[ semi-open interval \citep{park2009maximum}. Moreover, setting a uniform prior would be equivalent to assuming a specific range of dissipation scale and we found that in doing so the prior bounds were systematically hit. The exponential distribution has also the advantage of giving equal probability to all decades, that is, the dissipation scale is equally likely to be found in the range $[1;10]$\,kpc and in $[10;100]$\,kpc. \\
\indent The resulting posterior distributions are shown in Figure \ref{contour_kdiss_expprior}. For clarity, we show the results in terms of lengths ($L_{inj, diss}$), rather than wavenumbers ($k_{inj, diss}$). The posterior distributions are plotted with the 1 and 2-$\sigma$ confidence regions, and the marginalized distributions show the 1-$\sigma$ confidence interval as a shaded region. The blue distribution represents the result obtained in the uniform exposure configuration and the red stands for the mixed exposure configuration. The input values are recovered within 1-$\sigma$ in the distributions. However, they are  not centered on the input values, displaying a slight bias in the recovery of the parameters, which we argue could be due to the sample variance affecting this particular realization. As expected from Fig.~\ref{VaryingSFs}, the parameters are strongly degenerated. This is due to the nature of our choice of observable $SF$, as this range of covered scales does not allow us to fully break the degeneracy on the shape of the structure function between the parameters. The normalization parameter in particular is strongly underestimated. This can again be explained by the strong degeneracy between the parameters and the very large range allowed for the slope of the spectrum.  

\begin{figure}[h]
    \centering 
    \includegraphics[width=9cm]{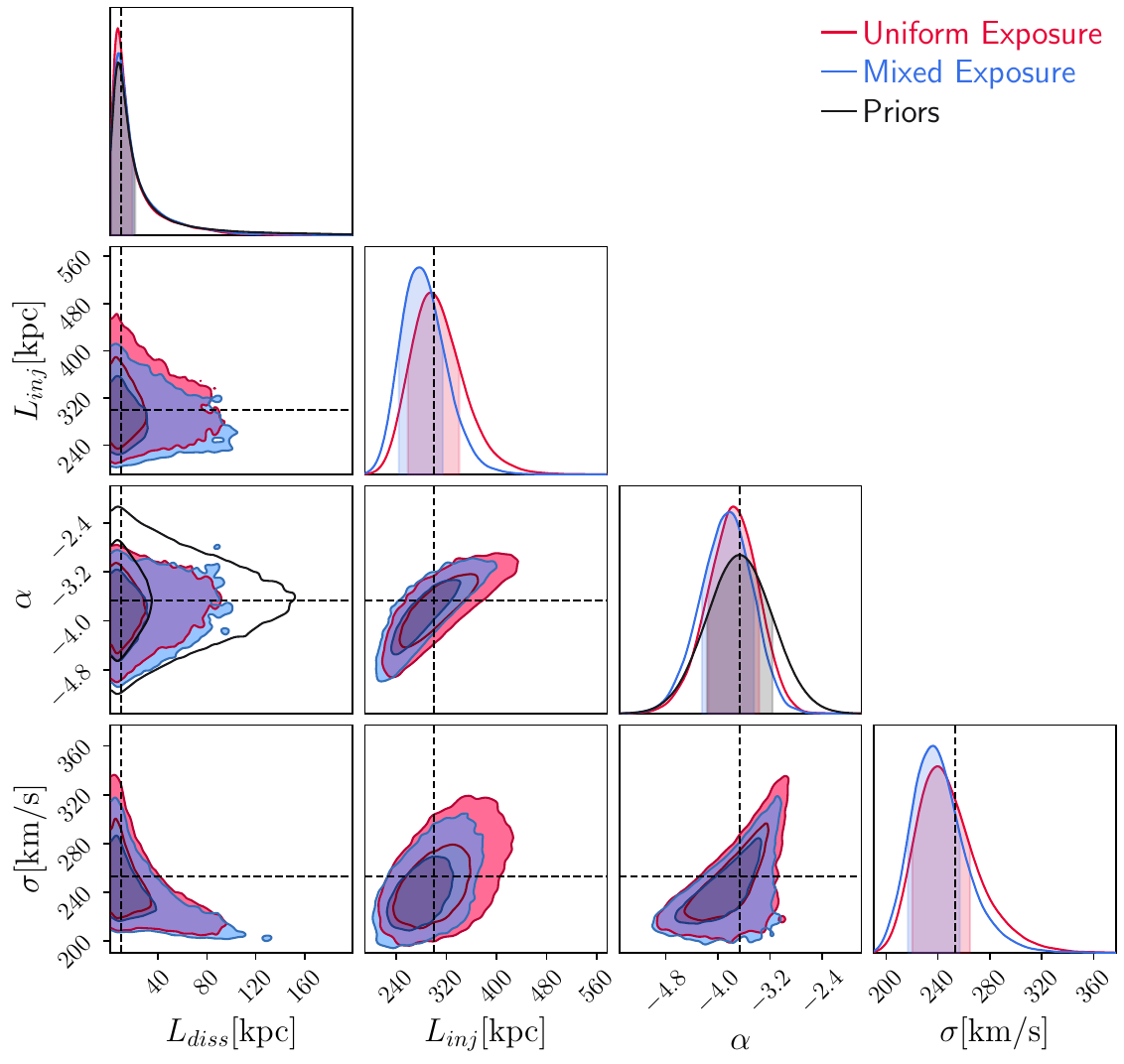}
    \caption{Similar to Fig.~\ref{contour_kdiss_expprior}, except that a Gaussian prior is set on parameter $\alpha$, instead of an uniform prior.}
    \label{contour_kdiss_exp_gauss_prior}
\end{figure}

\indent One could argue that a uniform prior on the slope of the power spectrum, especially a wide one, is a naive expectation, and could call for a more informative prior, on the basis of simulations or previous observations. For instance, \cite{Zhuravleva2012} used a dimensional analysis to argue that $-5<\alpha<-3$. Figure~\ref{contour_kdiss_exp_gauss_prior} shows the result of the same sampling procedure, however, with 
$\pi(\alpha) \sim \mathcal{N}(-11/3,0.5)$. This is a translation of the assumption that it is unlikely to find a turbulent power spectrum with a slope too far from that of a Kolmogorov cascade. In such a scenario, the prior of the slope is very informative or, conversely, the data are very uninformative with respect to this prior. This is again a translation of the difficulty of constraining this parameter. This results mainly in a slightly tighter distribution for the injection scale, as shown by the marginalized uncertainties shown in Table \ref{tab:model_params}. Adding a Gaussian prior to the slope of the power spectrum is only one option and one could equivalently do the same for the other parameters on the same assumptions of prior belief and/or knowledge.

\begin{table*}\renewcommand{\arraystretch}{1.8}
    \centering
    \caption{Summary table of the marginalized parameter uncertainties for each of the result figures presented in this section. The uncertainties given are $\pm 1 \sigma$. The first column refers to the model and its corresponding figure. The second and third column show the priors used for each configuration.}
    \label{tab:model_params}
    \begin{tabular}{ccccccc}
        \hline
                Model & $\pi(1/L_{diss})$ & $\pi(\alpha)$ & $L_{diss} [\mathrm{kpc}]$ & $L_{inj}  [\mathrm{kpc}]$ & $\alpha$ & $\sigma  [\mathrm{km}/\mathrm{s}]$ \\
 True Value& -& -& 10 & 300& $-11/3$ & 253\\ 
                \hline
                Uniform exp. (Fig. \ref{contour_kdiss_expprior})  & Exp($\frac{1}{10}$) & -- & $6.2^{+12.2}_{-4.7}$ & $244^{+73}_{-74}$ & $-3.98^{+0.86}_{-1.64}$ & $209^{+30}_{-17}$ \\
                Mixed exp. (Fig. \ref{contour_kdiss_expprior}) & Exp($\frac{1}{10}$) & -- & $6.4^{+12.4}_{-5.0}$ & $215^{+37}_{-75}$ & $-4.7^{+1.2}_{-2.1}$ & $200^{+21.5}_{-9.2}$ \\ 
        \hline
        Uniform exp. (Fig. \ref{contour_kdiss_exp_gauss_prior}) & Exp($\frac{1}{10}$) & $\mathcal{N}(-\frac{11}{3}, 0.5)$ & $6.6^{+12.0}_{-5.0}$ & $296^{+48}_{-38}$ & $-3.84^{+0.41}_{-0.40}$ & $233^{+26}_{-18}$ \\ 
                Mixed exp. (Fig. \ref{contour_kdiss_exp_gauss_prior}) & Exp($\frac{1}{10}$) & $\mathcal{N}(-\frac{11}{3}, 0.5)$ & $6.2^{+13.6}_{-4.9}$ & $275^{+39}_{-31}$ & $-3.91^{+0.37}_{-0.42}$ & $231^{+21}_{-17}$ \\ 
                \hline
        Uniform exp. (Fig. \ref{contour_gaussprior}) & -- & $\mathcal{N}(-\frac{11}{3}, 0.5)$ & -- & $301^{+48}_{-40}$ & $-3.78^{+0.37}_{-0.45}$ & $241^{+24}_{-17}$ \\ 
                Mixed exp. (Fig \ref{contour_gaussprior}) & -- & $\mathcal{N}(-\frac{11}{3}, 0.5)$& -- & $277^{+38}_{-31}$ & $-3.89^{+0.35}_{-0.43}$ & $238^{+19}_{-16}$ \\ 
        \hline
    \end{tabular}
\end{table*}

\begin{figure}[h]
    \centering 
    \includegraphics[width=9cm]{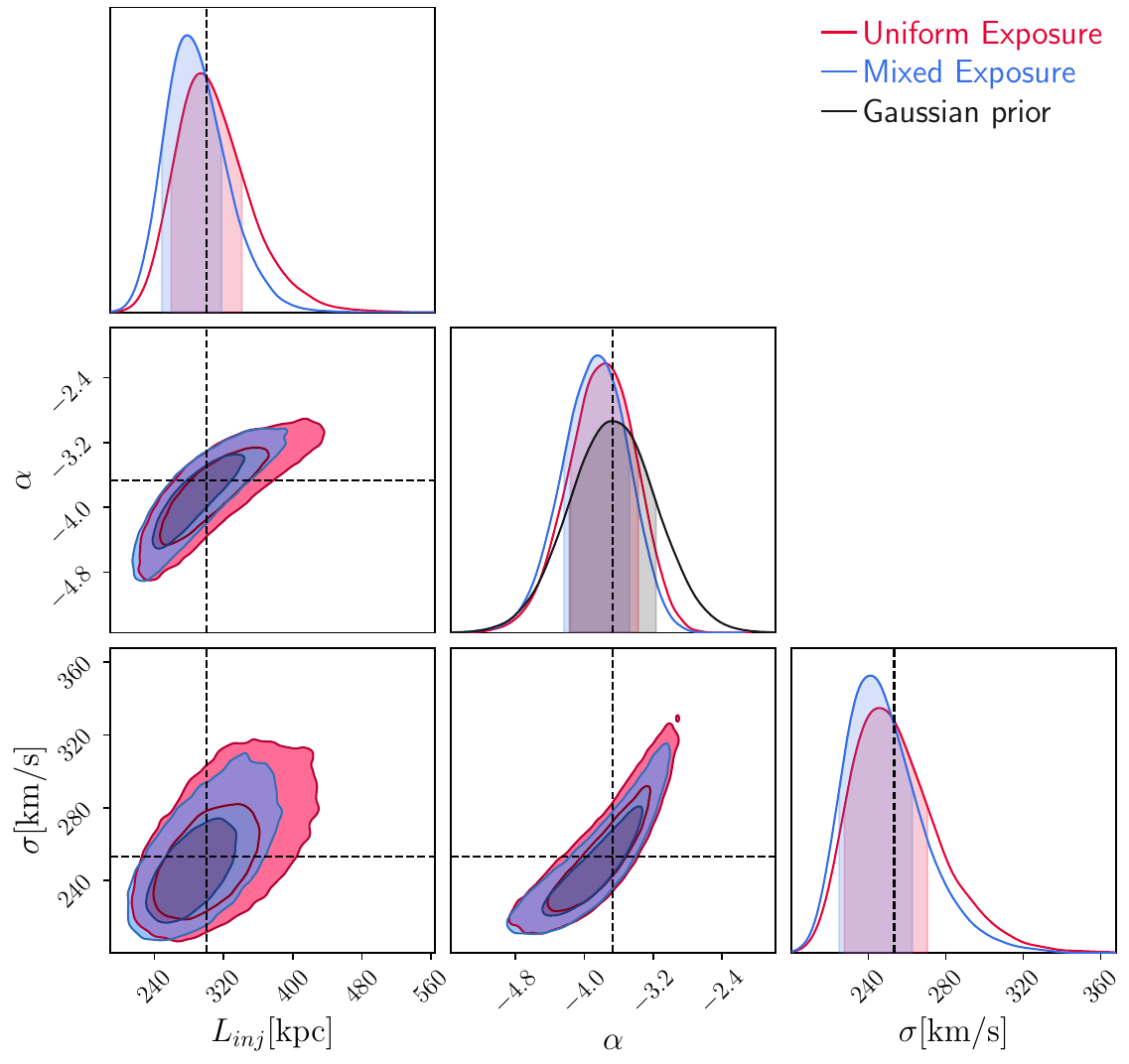}
    \caption{Posterior distribution obtained with setting the dissipation scale to the input value and letting the injection scale, norm and slope free. A Gaussian prior was set on the slope with $\pi(\alpha) \sim \mathcal{N}(-11/3,0.5)$, whereas the other parameters had uniform priors. The blue contour show the posterior distribution for the Mix exposure observation and the red for the uniform exposure. The prior is plotted in black. Dotted lines show the input values of the simulations.}
    \label{contour_gaussprior}
\end{figure}

\indent As the influence of the dissipation scale on the SF shape is very limited on the range of scales we investigate here, it is reasonable to set the dissipation scale to the simulation input value. The result of this is shown in Fig.~\ref{contour_gaussprior}. The Gaussian prior was kept on the slope parameter to restrict its range to the most reasonable values. The marginalized uncertainties of all our MCMC procedures are shown in Table \ref{tab:model_params}. The resulting uncertainties are not significantly better than in a configuration where the dissipation scale is left free with an exponential prior. This is due to the prior being strongly constraining on the dissipation scale, and its aforementioned weak dependency on the SF.

\section{Discussion}
\label{s:disc}

\indent We have accounted for the total error budget in our analysis of mock observations of a galaxy cluster with the X-IFU instrument, including the sample variance. To do so, we have formulated  simplifying assumptions in the computation of the errors on the structure function. As such, our total error budget computation is an approximation. It is likely that the full computation of the error would lead to varying results. However, it is unlikely that it would significantly alter the precision on the recovery of the turbulent parameters. \\
\indent We purposely chose not to include any astrophysical background in our mock observations. The addition of astrophysical background should, in principle, only increase the uncertainty in the spectral fitting and hence on the recovered velocities. As such, it should translate into an increase of $\sigma_{stat}$, thus marginally affecting the nature of our results. The main impact should be on the width of the posterior distributions of the turbulent parameters which would be larger, as a result of the larger error. In addition, with our current exposure and signal-to-noise binning setup, the background is expected to be subdominant over all observed regions and should not influence  the line velocity measurements significantly \citep{cucchetti2018reproducibility}.  The astrophysical background will be included in a forthcoming study investigating the optimal realistic observing strategy.  \\
\indent Our correction of the structure function and computation of the error assumes a perfect knowledge of the statistical error, $\sigma_{stat}$ (see Sect.~\ref{sig_stat}). Quantifying this term could in principle be done with bootstrapping from observations. The possible error on the estimation would have to be properly propagated to the final distribution of the parameters. Equivalently, we could make use of complete simulations to assess $\sigma_{stat}$ given an observational setup. \\
%
\indent Our toy model cluster is obviously not plainly representative of real clusters. In a complete analysis, modeling the structure function would have to take into account the departure to the spherical symmetry of all physical quantities governing the emissivity of the ICM. Same goes for the underlying turbulent field, that was assumed to be isotropic and homogeneous. This is not likely to be the case in real clusters, given the complex dynamics and substructures altering the turbulent velocity field, as well as the gravitational field breaking isotropy. \\
%
\indent In addition to the previous comments, we should emphasize the other possible contributions to the velocity field within galaxy clusters. Indeed, bulk motion within the cluster, or gravitational redshift gradients \citep{2023A&A...679A..24M} will be measured through the centroid shift as well and will bias the structure function and the subsequent turbulent parameters. Their effect will need careful handling and comparison with cosmological simulations.  \\
\indent The numerical implementation of the analytic model requires a number of simplifying assumptions to adapt the formalism to the non-ideal situation of a realistic cluster profile with irregular regions. This pushes the edge of using analytical treatment in the prediction of such observables. In a forthcoming study, we shall investigate methods such as simulation based inference (SBI, \cite{tejero-cantero2020sbi}), which requires a great number of simulations, to derive a full error budget without any approximations. It would also naturally solve the issue of the inclusion of the structure function's covariance. This would allow us to include other observables, such as the line broadening, which was not included in the present study, as we deliberately focused on the velocity structure function. 
Such added observables would help break the degeneracy between the turbulent parameters.

\section{Conclusions}
\label{s:conc}

With this study, we have further investigated the ability to characterize the turbulent motions in galaxy clusters with integral field spectroscopy in the X-ray domain. We developed a framework incorporating a comprehensive error budget and we applied it to mock observations (with unrealistic exposure time) of a case study cluster with  the X-IFU instrument on board the Athena mission. 
\begin{itemize}
    \item Using end-to-end simulations (on the one hand) and modeling of structure functions for a physically sound cluster model (on the other), we performed an MCMC sampling with a semi-analytical model to the structure function in order to retrieve the parameters from a simple Kolmogorov-like turbulent model (injection scale, slope, and normalization factor i.e., velocity dispersion).
    \item The error model presented here relies on a careful account of all errors terms \citep[i.e. the sample variance and additional statistical terms as defined in][]{Cucchetti2019}, and on the further development of the formalism to compute the structure function and errors from \citet{Clerc2019}. Such a comprehensive account is mandatory in any analysis aiming at a proper physical characterisation of the ICM turbulence. As a drawback, it affects the final constraints on the turbulence parameters. Overcoming the remaining assumptions made for the error computations is the goal to reach in order to plainly assess the full potential of the future  XIFU observations to characterize the physical properties of the ICM turbulence. This will constitute the basis for future work.
    \item Under the observational conditions of this paper, the constraints on turbulent parameters are strongly degenerate. Breaking such degeneracies needs either prior knowledge of some of the parameters or different strategies to access a wider range of spatial scales with sufficiently high resolution and S/N values. Solving this problem likely requires optimizing the strategy by exploring a relatively wide configuration space. 

\item In the most favorable observational configuration and assuming prior knowledge of the dissipation scale and power spectrum slope, we were able to constrain the injection scale to +16\%/-13\%, velocity dispersion to +10\%/-7\%, and slope to +10\%/-12\% of the input values. 
    
\end{itemize}

In the upcoming  work in this series of papers, we will look into the optimal observing strategy. We will define the best combination of X-IFU pointings, in terms of geometrical configuration and exposure time to extract the most stringent constraints on the turbulent parameters. We will also address the possibility of stacking multiple observed clusters, taking into consideration the stochastic nature of the turbulent process. Co-adding independent observations of clusters together should reduce the errors on each parameter by a factor of $\sqrt{N}$, with N the number of clusters \citep[see e.g.,][]{Dupourque2023}.

Finally, we stress that this study was conducted with the Athena/X-IFU mission configuration before the mission reformulation by the European Space Agency\footnote{ESA just rendered public the new specification of the Athena mission in november 2023.}. The newly issued specifications of the \emph{Athena} mission affect top-level requirements, such as the effective area, spectral resolution, or X-instrument field-of-view. These changes will likely  quantitatively impact the results presented here. Our future work will focus on  and integrate these new instrumental specifications.

\begin{acknowledgements}
SB, AM, EP and NC acknowledge the support of CNRS/INSU and CNES. The following python packages have been used throughout this work : \texttt{astropy} \citep{astropy:2013, astropy:2018, astropy:2022}, \texttt{chainconsumer} \citep{Hinton2016}, \texttt{emcee} \citep{2013PASP..125..306F}, \texttt{matplotlib} \citep{Hunter:2007} and \texttt{cmasher} \citep{2020JOSS....5.2004V}. 
\end{acknowledgements}

\bibliographystyle{aa}
\bibliography{biblio_2}

\appendix


\section{Modeling $\meansf$ in arbitrary spatial binning configurations\label{app:sfgrid}}

The model for the expectation value $\meansf$ of the structure function $SF$ requires knowledge of the underlying turbulent velocity field and the emissivity field. Both these quantities are defined on a 3D spatial grid $(x, y, z) = (x, \vec\theta)$. An additional important input is the geometrical design describing the position and shape of 2D bins within which X-ray spectra are extracted and from which a line centroid shift is measured.
The formalism of \citet{Clerc2019} does not provide a closed formula for $\meansf$ in the very general case of arbitrary bin shapes and arbitrary emissivity field.
This section intends to extend their approach with this purpose in mind; we derive $\meansf$ in a form that relies on pre-computed Fourier transform of the emissivity field.

We start from the definition of the structure function $SF(\sepa)$ computed from line centroid shifts $C_{\ww}$, each associated to a spatial bin $\mathcal{W}$:
\begin{equation*}
SF(\sepa) = \frac{1}{N_p(\sepa)} \sum_{d(\ww, \wwp) = s} \left | C_{\wwp} - C_{\ww} \right | ^2.
\end{equation*}
In this formula, the sum runs over all pairs of bins separated by a distance $s$, the total number of them reads $N_p(\sepa)$.
Given a 2D geometry, the list of pairs and of pairwise distances is known and it is a fixed input in the problem.

Developing this expression and averaging over many realizations of the turbulent velocity field, we are left with computing $\langle C_{\ww} C_{\wwp} \rangle$ for any two bins $\ww$ and $\wwp$ (including the case of $\ww = \wwp$). We write the line centroid shift in bin $\ww$ as a flux-weighted average of centroid shifts projected along the \AM{LoS} (i.e.,~along coordinate $x$):

\begin{align*}
    C_{\ww} & = F_{\ww}^{-1} \int  \dd \vec\theta \ww(\vec\theta) \int \dd x \epsilon(x, \vec\theta) v(x, \vec\theta)  \\
            & = F_{\ww}^{-1} \int  \dd \vec\theta \ww(\vec\theta) F(\vec\theta) C(\vec\theta),
\end{align*}
where $v(x, \vec\theta)$ stands for the \AM{LoS} velocity, $\epsilon$ is the (3D) emissivity, $F(\vec\theta)$ is the line flux along a single \AM{LoS} labeled by the sky coordinate, $\vec \theta,$ and $F_{\ww}$ is the line flux summed within bin, $\ww$.
Decomposing the velocity field into its Fourier components, $V_{\kthree}$, we obtain an expression for the velocity shift along a single \AM{LoS} direction:
\begin{equation}
    \label{eq:appendix_c_theta}
    C(\vec\theta) = \sum_{\kthree} V_{\kthree} e^{i \omega \vec \xi \cdot \vec \theta} \int  e^{i\omega k_x x} \rho(x, \vec\theta) \dd x
,\end{equation}
where $\kthree = (k_x, \vec \xi)$ is the Fourier coordinate, $\omega = 2\pi/L$, $L$ is a large characteristic length and $\rho(x, \vec\theta) \equiv \epsilon(x, \vec\theta)/F(\vec\theta)$.

Using the definition of the velocity power-spectrum $P_{3D}$ and writing complex conjugation with symbol ${}^*$, we find:
\begin{equation*}
    \langle C_{\ww} C_{\wwp} \rangle = F_{\ww}^{-1} F_{\wwp}^{-1} \sum_{\kthree} P_{3D}(\kthree) c_{\epsilon \cdot \ww}(\kthree) c_{\epsilon \cdot \wwp}^*(\kthree).
\end{equation*}
The complex quantity $c_{\epsilon \cdot \ww}(\kthree)$ depends solely on the bin geometry, size, position and on the emissivity. Its value can be pre-calculated as the (3D) Fourier transform of the quantity $(\epsilon \cdot \ww)$, namely:
\begin{equation*}
c_{\epsilon \cdot \ww}(\kthree) = \int e^{i \omega \kthree \cdot \xthree} \ww(\vec\theta) F(\vec\theta) \rho(\xthree) \dd \xthree,
\end{equation*}
where the function $\ww(\vec\theta)$ takes value 1 within bin $\ww$ and zero outside of it.
We may then reassemble the terms and use the fact that $SF$ is a real quantity to obtain the formula shown in Eq.~\ref{eq:sf_grid_fourier}.

While varying the characteristics of the turbulent field, the term between parentheses in Eq.~\ref{eq:sf_grid_fourier} remains constant. Therefore it can be computed only once, which significantly accelerates the exploration of the turbulent parameters through MCMC sampling. 
The main burden then resides in pre-computing $c_{\epsilon \cdot \ww}$ for each bin $\ww$. 
In general\footnote{The case of $\beta$-model emissivity fields benefits from a number analytical expressions that may accelerate the process, see App.~B in \citet{Clerc2019}.}, this calculation requires numerical evaluations of the (3D) Fourier transforms of the field $(\epsilon \cdot \ww)$.
These can be simplified and accelerated if the emissivity is constant within a bin $\ww$, regardless of the \AM{LoS} $\vec\theta$, i.e.~if $\epsilon(x, \vec\theta) = \varepsilon(x)$ and $F(\vec\theta) = F$. This is a relevant approximation if the bins are small in comparison to typical emissivity gradients. There, we obtain:
\begin{align*}
    c_{\epsilon \cdot \ww}(k_x, \vec\xi) & \simeq \widetilde{\varepsilon}(k_x) \times \widehat{\ww}(\vec \xi), \\
F_{\ww} & \simeq F \times S_{\ww.}
\end{align*}
In these expressions, $S_{\ww}$ is the surface area of a bin, tilde and hat stand for 1D and 2D Fourier transforms respectively. In this paper, we make use of this approximation in order to make calculations tractable.

\section{Computation of $\sigma_D^2 = {\rm Var}(D)$}
\label{app:vard}

\begin{figure}
    \centering
    \includegraphics[width=\linewidth]{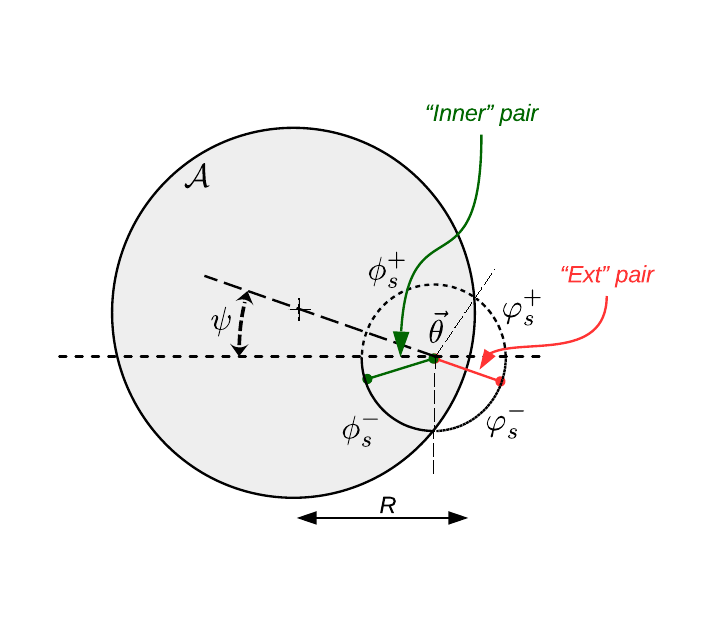}
    \caption{Geometrical quantities used in counting `ext' pairs of length $\sepa$ from a point $\vec\theta$ located within the circular region of analysis, $\mathcal{A,}$ of radius, $R$, at a distance $\theta = |\vec\theta|$ from the center. Whether pairs are pointing northward or southward, their number depends on the angle $\varphi^{+}$ or $\varphi^{-}$ and we have $\varphi=\varphi^+ + \varphi^-$. Symmetrical relationships arise for the number of `inner' pairs, with angles $\phi^{+}$ and $\phi^{-}$.}
    \label{fig:appendix_vard_geometry}
\end{figure}

\citet{Cucchetti2019} introduced the quantity $D(\sepa)$ whose variance intervenes in the calculation of uncertainties of the structure function. They defined:
\begin{equation}\label{eq:d_definition}
D(\sepa) = \frac{1}{N^{in}(\sepa)} \sum_{in} \left[ C(\vec \theta) - C(\vec \theta^{\prime}) \right]
.\end{equation}
Similarly as in Appendix~\ref{app:sfgrid}, $C(\vec\theta)$ represents the line centroid shift along a \AM{LoS} labeled by the sky coordinate $\vec\theta$. We arbitrarily set $C(\vec\theta) = 0$ for locations outside of the region of analysis.
The sum runs over all pairs of coordinates $(\vec\theta, \vec\theta^{\prime})$ separated on sky by a distance, $s$, both being located inside the region of analysis (hence, the label "in"). In the following, we assume the region of analysis is circular of radius, $R$. The number of such pairs is $N^{in}$ and its formal derivation is provided in Appendix~D of \citet{Clerc2019}. The above definition for $D$ is however ambiguous and it hints to a necessary revision of the \citet{Cucchetti2019} formula, which is out of the scope of the present study. We disambiguate the expression by requiring that the pairs under the sum sign are ordered such that $\vec\theta$ always lies southwards of $\vec\theta^{\prime}$. The following calculation introduces an extra set of pairs separated by a distance $\sepa$, labeled "ext," which cross the border of the region of analysis (see Fig.~\ref{fig:appendix_vard_geometry}).
We first compute the sum over all the pairs (i.e., both the~"in" and "ext" pairs) and we then subtract an explicit derivation of the contribution from the "ext" pairs.

Let us introduce the new quantity $D^{\prime}$ through:
\begin{equation*}
    D^{\prime}(\sepa) = \frac{1}{N^{in}(\sepa)} \sum_{in+ext} \left[ C(\vec \theta) - C(\vec \theta^{\prime}) \right],
\end{equation*}
in which the sum runs over all pairs (always assuming the conventional orientation toward the north).
Rearranging the summation terms and using that $C(\vec\theta)=0$ outside of the region of analysis, we find $D^{\prime}(\sepa)=0$.

We are thus left with calculating the sum over all pairs of kind `ext'. Those pairs having their southern end $\vec\theta$ within the analysis region contribute a term $C(\vec\theta)$; while those having their southern end outside of the region of analysis contribute a term $-C(\vec\theta^{\prime})$.
To each direction, $\vec\theta,$ within the field of analysis, $\mathcal{A,}$ we associate the number of "ext" pairs: $\varphi_{\sepa}(\vec \theta)$. This number can be decomposed into pairs that point northwards ($\varphi^{+}_{\sepa}$) and pairs that point southwards ($\varphi^{-}_{\sepa}$). We write:

\begin{equation*}
\sum_{ext} \left[ C(\vec \theta) - C(\vec \theta^{\prime}) \right] = \sum_{\vec \theta \in \mathcal{A}} \left( \varphi^{+}_{\sepa}(\vec \theta) -\varphi^{-}_{\sepa}(\vec \theta) \right) C(\vec \theta)
\end{equation*}

Introducing the number of "in" pairs $\phi_{\sepa}(\vec \theta) = 2\pi - \varphi_{\sepa}(\vec\theta)$, it is clear that the following relations hold (e.g., see Fig.~\ref{fig:appendix_vard_geometry}):
\begin{eqnarray*}
\varphi_{\sepa} =  \varphi^{+}_{\sepa} + \varphi^{-}_{\sepa} & ; &
\phi_{\sepa}  =   \phi_{\sepa}^{+} + \phi_{\sepa}^{-}; \\
\phi_{\sepa}^{+}  =  \pi - \varphi_{\sepa}^{+} & ; &
\phi_{\sepa}^{-}  =  \pi - \varphi_{\sepa}^{-}.
\end{eqnarray*}
An explicit expression for $\phi_{\sepa}$ is given in Appendix~D of \citet{Clerc2019}. Hence, we only need to calculate $\phi_{\sepa}^{+}$ from which we will deduce the other quantities ($\varphi_{\sepa}^{-}$, $\varphi_{\sepa}^{+}$, and $\phi_{\sepa}^{-}$).

\begin{figure}
    \centering
                \includegraphics[width=\linewidth]{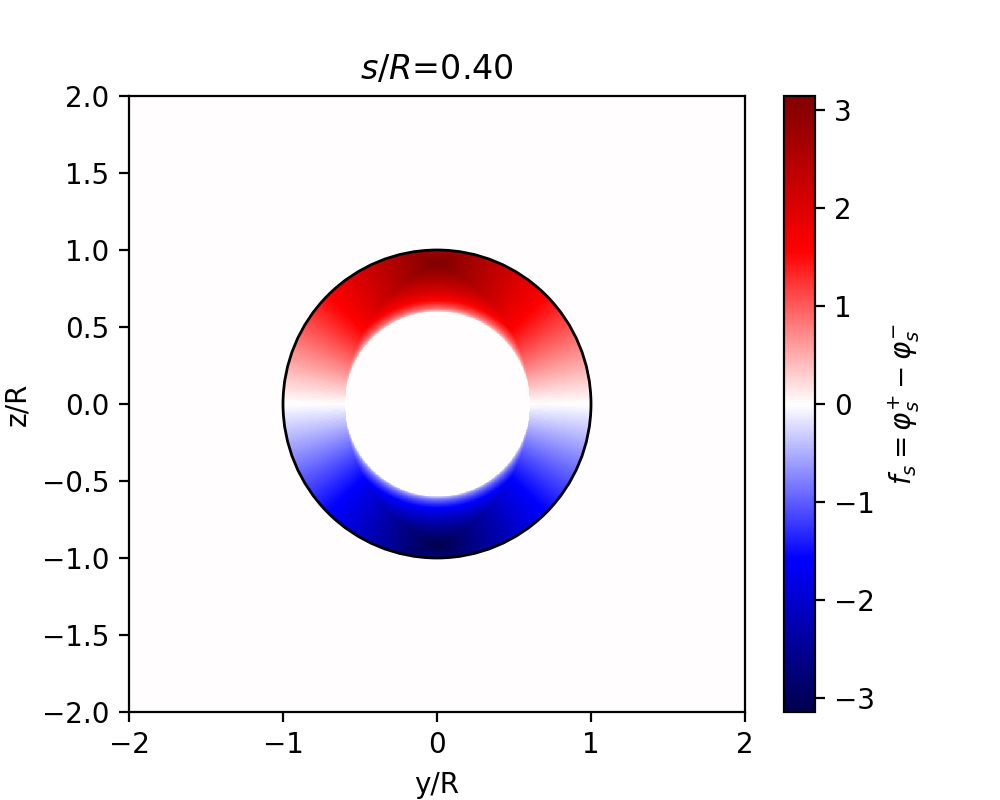}
    \caption{Representation of the functional $f_s(\vec\theta)$. It takes value in $[-\pi, \pi]$ for each point $\vec\theta=(y, z)$ located within the circular region of analysis. This region of radius, $R,$ is bounded by the plain black line.}
    \label{fig:appendix_f_function}
\end{figure}

Let us describe the position of the point $\vec\theta$ by two coordinates: $\theta = |\vec\theta|$ its distance to the center of the circular shape of $\mathcal{A}$~; and $\psi$ is the angle $\vec\theta$ makes with the horizontal line. Geometrical considerations lead to distinguish five mutually exclusive cases:
\begin{equation*}
\phi_{\sepa}^{+}(\theta, \psi) = \left\{
        \begin{array}{cl}
        \pi                                     & {\rm if\ } \tau_{\sepa}(\theta, \psi) > \pi, \\
        \phi_{\sepa}(\theta) - \pi      & {\rm if\ } \tau_{\sepa}(\theta, \psi) < \phi_{\sepa}(\theta) - \pi, \\
        0                                               & {\rm if\ } \tau_{\sepa}(\theta, \psi) < 0, \\
        \phi_{\sepa}(\theta)    & {\rm if\ } \tau_{\sepa}(\theta, \psi) > \phi_{\sepa}(\theta), \\
        \tau_s(\theta, \psi)    & \mathrm{otherwise,} \\
        \end{array}\right.
\end{equation*}
where $\tau_{\sepa} (\theta, \psi) = \phi_{\sepa}(\theta)/2 - \psi$ and $\tau_{\sepa} \in [-\pi/2, 3\pi/2]$.
Figure~\ref{fig:appendix_f_function} shows one example ($s=0.4R$) representation of the function $f_s(\vec\theta) = \varphi^{+}_{\sepa}(\vec \theta) -\varphi^{-}_{\sepa}(\vec \theta)$.
This purely geometrical function associates to each point within the field of analysis the balance between the number of "ext" pairs pointing northwards and those pointing southwards.

Using that $D^{\prime}=0$, we write:
\begin{equation*}
    D(\sepa) = - \frac{1}{N^{in}(\sepa)} \sum_{\theta \in \mathcal{A}} f_s(\vec \theta) C(\vec \theta).
\end{equation*}

Since $\langle C \rangle = 0$, we have that $\langle D \rangle = 0$ and $Var(D)=\langle D^2 \rangle$.
Developing the calculation involves the quantity $\left \langle C(\vec \theta_1) C(\vec \theta_2) \right\rangle$ and Eq.~\ref{eq:appendix_c_theta} may be used to obtain:
\begin{equation*}
    Var(D) = \frac{1}{(N^{in})^2} \sum_{\kthree} P_{3D}(k) \left| \int_{\vec \theta \in \mathcal{A}} \dd \vec \theta f_s(\vec \theta) e^{i \omega \vec \xi \cdot \vec \theta} \widetilde{\chi^{\vec \theta}}(k_x) \right|^2.
\end{equation*}
The function $\widetilde{\chi^{\vec\theta}}(k_x)$ may be considered either as the (1D) Fourier transform of the function $\chi^{\vec\theta}(x) = \rho(x, \vec\theta)$; or, if seen as a function of $\vec\theta$, it is the (2D) inverse Fourier transform of the function $u_{k_x}(\vec\xi) = \widetilde{\rho}(k_x, \vec\xi)$ (up to a constant factor). Hence, we obtain the following numerically tractable expression:
\begin{equation}
    \label{eq:appendix_varD}
    Var[D(\sepa)] = \frac{1}{N^{in}(\sepa)^2} \left( \frac{\omega}{2 \pi} \right)^4 \sum_{\kthree} P_{3D}(k) \left|  \left( u_{k_x} \otimes \widehat{f_{\sepa}} \right) (\vec \xi) \right|^2
,\end{equation}
involving a series of 2D convolutions indicated with the sign $\otimes$. We remind that a hat indicates 2D Fourier transform.

In practical applications, the computation of $D(s)$ is performed in spatial bins of finite size. The derivation above assumes infinitely small bins. Similarly as in \citet[][in particula, Appendix~E]{Clerc2019}, we introduce an approximate scheme for dealing with spatial bins of finite size, $\ell$, all supposedly of identical shape and regularly spaced. This approximation replaces $C(\vec\theta)$ above by a filtered version:
\begin{equation*}
    C(\vec \theta) \rightarrow \int \dd \vec \mu \mathcal{F}_{\ell}(\vec \mu) C(\vec \theta -\vec \mu),
\end{equation*}
where $\mathcal{F}_{\ell}$ is a top-hat convolution filter describing the shape of bins. This approximation works under the assumption of sufficiently small bins with respect to the size of the analysis bin and to typical variations of the emissivity. The corrected equation for ${\rm Var}(D)$ is obtained by replacing $\widehat{f_{\sepa}}$ by the product $\widehat{f_{\sepa}}\widehat{\mathcal{F}_{\ell}}^*$ in Eq.~\ref{eq:appendix_varD}.

\section{Plane-constant emissivity model}
\label{app:xbeta}

An useful simplified emissivity model consists of a $\beta$-model along the \AM{LoS} direction $x$, identical for any selected \AM{LoS} $\vec \theta$. We call this model Xbeta, it is given by:

\begin{equation}
    \label{eq:appendix_xbeta}
    \epsilon(x, \vec \theta) = \varepsilon(x) = \varepsilon(0)\left( 1+ \frac{x^2}{r_c^2 + \theta_0^2} \right)^{-3 \beta}
.\end{equation}

This expression involves a core-radius, $r_c$, and a fixed impact parameter, $\theta_0$. The characteristic size of the model along the \AM{LoS} direction is $R_c = (r_c^2+\theta_0^2)^{1/2}$. This emissivity model does not depend on the sky coordinate $\vec \theta$, hence, it produces uniform surface brightness. Analytical expressions for its Fourier transform make the computation of the 2D power spectrum of the centroid shift easily tractable \citep{Churazov2012, ZuHone2016, Clerc2019}.

\section{Approximate scaling of sample variance}
\label{app:sf_scalingerr}

This section details the calculation leading to approximate scalings of the sample variance in the structure function.
To this end, we restrict the following discussion to emissivity field described by Eq.~\ref{eq:appendix_xbeta}, regularly binned centroid shift maps and a very large region of analysis. Under these assumptions, we have \citep[see Eq.~12 and~13 in][]{Clerc2019}:

\begin{equation}\label{eq:appendix_sf_mean}
{\rm sf}(\sepa) = 4 \pi \left(\frac{\omega}{2 \pi}\right)^2 \int_{0}^{\infty} P_{\ell} P_{2D}^{\infty} \left(1 - J_0(\omega \xi \sepa) \right) \xi \dd \xi
,\end{equation}

\begin{equation}\label{eq:appendix_sf_var}
{\sigma_{\rm sf}^2}(\sepa) = \frac{16 \pi}{\fovarea} \left(\frac{\omega}{2 \pi}\right)^2 \int_{0}^{\infty} P_{\ell}^2 {P_{2D}^{\infty}}^2 \left(1 - J_0(\omega \xi \sepa) \right)^2 \xi \dd \xi
,\end{equation}

where ${\rm sf} = \langle SF \rangle$, $\sigma_{\rm sf}^2 = \langle (SF- {\rm sf})^2 \rangle$. In these expressions, $\omega = 2 \pi/L$ with $L$ an arbitrary length setting the length unit system~; $P_{\ell}(\xi)$ is the power spectrum of the 2D bins, each of characteristic size, $\ell$~; $P_{2D}^{\infty}(\xi)$ is the power spectrum of the centroid shift map, extrapolated to an unbounded region of analysis; $\fovarea$ denotes the area of the region of analysis; $J_0$ is the Bessel function of the first kind of order 0. Furthermore, we specify the emissivity model following Eq.~\ref{eq:appendix_xbeta}, involving a characteristic size, $R_c$.

In what follows we will be interested in the relative sample variance, as given by $\sigma_{\rm sf}/{\rm sf}$.
We formulate a number of additional assumptions in order to derive useful approximations:
\begin{itemize}
    \item The 3D velocity power spectrum is a power-law on the interval bounded by the injection and dissipation scales and vanishes outside of this interval. We note this is a simplification of the actual power-spectrum used in our study (Eq.~\ref{P3Deqn}).
    \item All spatial bins are of identical size $\ell$ and the Fourier transform of their shape is well approximated by a top-hat function over the interval: $-1/\ell < \omega \xi < 1/\ell$.
    \item Bins are small with respect to the projected injection length and larger than the dissipation length of turbulent motions: $k_{\rm inj} < 1/\ell < k_{\rm diss}$.
\end{itemize}

The expression for $P_{2D}^{\infty}$ provided in Eq.~C.4 of \citet{Clerc2019} makes evident the role played by the emissivity profile as a low-pass filter, with characteristic cut-off frequency $1/R_c$:
\begin{equation}
\label{eq:appendix_p2d_infinity}
P_{2D}^{\infty}(\xi) \propto \int P_{3D}\left( \sqrt{k_x^2 + \xi^2} \right) P_{\varepsilon}(k_x) \dd k_x
,\end{equation}
where $P_{\varepsilon}$ represents the 1D power spectrum of the emissivity profile. Two distinct cases appear, according to $R_c$ being larger (case A) or smaller (case B) than the turbulent injection scale.

    \subsection{Study of case A ($1/R_c < k_{inj}$)}
In case A, the term under the integral in Eq.~\ref{eq:appendix_p2d_infinity} contributes non-zero terms only for values of $k_x$ very close to zero. We then have $P_{2D}^{\infty}(\xi) \propto P_{3D}(\xi)$ and we approximate the result by writing:
\begin{equation*}
P_{2D}^{\infty}(\xi) \simeq \left\lbrace
        \begin{aligned}
                A \left(\frac{\omega \xi}{k_{inj}}\right)^{\alpha} &{\ \ \ } {\rm if\ } \omega\xi \in [k_{inj}; k_{diss}], \\
                0 &{\ \ \ }{\rm otherwise.}
        \end{aligned}
        \right.
\end{equation*}
Here, $A$ is a constant that cancels out in forming the ratio $\sigma_{\rm sf}/{\rm sf}$. The integrals involved in expressions~\ref{eq:appendix_sf_mean} and~\ref{eq:appendix_sf_var} extend from $0$ to $+\infty$. The expressions for $P_{\ell}$ and $P_{2D}$ show it is only necessary to compute the integrals over the domain $[k_{inj}, 1/\ell]$. Two regimes are then worth investigating.

At small separations $\sepa \ll \ell$, the second-order Taylor expansion of the Bessel function $J_0$ leads to:
\begin{equation}\label{eq:appendix_invsnr_small_A}
\frac{\sigma_{\rm sf}}{\rm{sf}} \simeq 2 \sqrt{\pi} \frac{L_{inj}}{\fovarea^{1/2}} \frac{\mathcal{G}\left(r, 2 \alpha+6 \right)^{1/2}}{\mathcal{G}\left(r, \alpha+4 \right)} {\ \ \ \ \ \ {\rm if\ } \sepa \ll \ell.}
\end{equation}
Here, we define $r = \ell k_{inj}  \in [0, 1]$ and the specific function:
\begin{equation*}
\mathcal{G}(r, n) = \frac{1}{n} \left( r^{-n} -1 \right).
\end{equation*}

At large separations, $\sepa \gg L_{inj}$, assuming that the Bessel function vanishes toward large values of its argument:
\begin{equation}\label{eq:appendix_invsnr_large_A}
\frac{\sigma_{\rm sf}}{\rm{sf}} \simeq 2 \sqrt{\pi} \frac{L_{inj}}{\fovarea^{1/2}} \frac{\mathcal{G}\left(r, 2 \alpha+2 \right)^{1/2}}{\mathcal{G}\left(r, \alpha+2 \right)} {\ \ \ \ \ \ {\rm if\ } \sepa \gg L_{inj}}
.\end{equation}

    \subsection{Study of case B ($k_{inj} < 1/R_c < k_{diss}$)}

In case B, the integral in Eq.~\ref{eq:appendix_p2d_infinity} decomposes into a sum over three intervals: $[0, k_{inj}]$, $[k_{inj}, 1/c]$ and $[1/R_c, +\infty]$. A close examination of the integrals for various values of $\xi$ leads to the following approximation:

\begin{equation*}
P_{2D}^{\infty}(\xi) \simeq \left\lbrace
        \begin{aligned}
                B &{\ \ \ } {\rm if\ } \omega\xi \ < k_{inj}, \\
                B \left(\frac{\omega \xi}{k_{inj}}\right)^{\alpha} &{\ \ \ } {\rm if\ } \omega \xi \in [k_{inj}; k_{diss}], \\
                0 &{\ \ \ }{\rm otherwise.}
        \end{aligned}
        \right.
\end{equation*}
Here, $B$ is a constant of no relevance in the discussion. The expressions for $P_{\ell}$ and $P_{2D}^{\infty}$ show that integrals~\ref{eq:appendix_sf_mean} and~\ref{eq:appendix_sf_var} must be calculated over domain $[0, 1/\ell]$. Similarly as in case A, we identify two limiting regimes, allowing for a Taylor expansion of the $J_0$ Bessel function under the integrals. It follows that:

\begin{equation}\label{eq:appendix_invsnr_small_B}
\frac{\sigma_{\rm sf}}{\rm{sf}} \simeq 2 \sqrt{\pi} \frac{L_{inj}}{\fovarea^{1/2}} \frac{\left[\frac{1}{6} + \mathcal{G}\left(r, 2 \alpha+6 \right)\right]^{1/2}}{\frac{1}{4} + \mathcal{G}\left(r, \alpha+4 \right)} {\ \ \ \ \ \ {\rm if\ } \sepa \ll \ell,}
\end{equation}

\begin{equation}\label{eq:appendix_invsnr_large_B}
\frac{\sigma_{\rm sf}}{\rm{sf}} \simeq 2 \sqrt{\pi} \frac{L_{inj}}{\fovarea^{1/2}} \frac{\left[\frac{1}{2} + \mathcal{G}\left(r, 2 \alpha+2 \right)\right]^{1/2}}{\frac{1}{2} + \mathcal{G}\left(r, \alpha+2 \right)} {\ \ \ \ \ \ {\rm if\ } \sepa \gg L_{inj}}
.\end{equation}

    \subsection{Interpretation}
Under the assumptions stated in this Appendix, the relative uncertainties due to sample variance scale as provided in Eq.~\ref{eq:appendix_invsnr_small_A}, \ref{eq:appendix_invsnr_large_A}, \ref{eq:appendix_invsnr_small_B}, and~\ref{eq:appendix_invsnr_large_B}, each corresponding to a limiting case obtained by comparing together the values of $R_c$ (typical size of the galaxy cluster), of $L_{inj}$, $L_{diss}$ (typical scales of the turbulent field) and of $\ell$ (typical size of the spatial binning scheme).

\begin{figure}
    \centering
                \includegraphics[width=\linewidth]{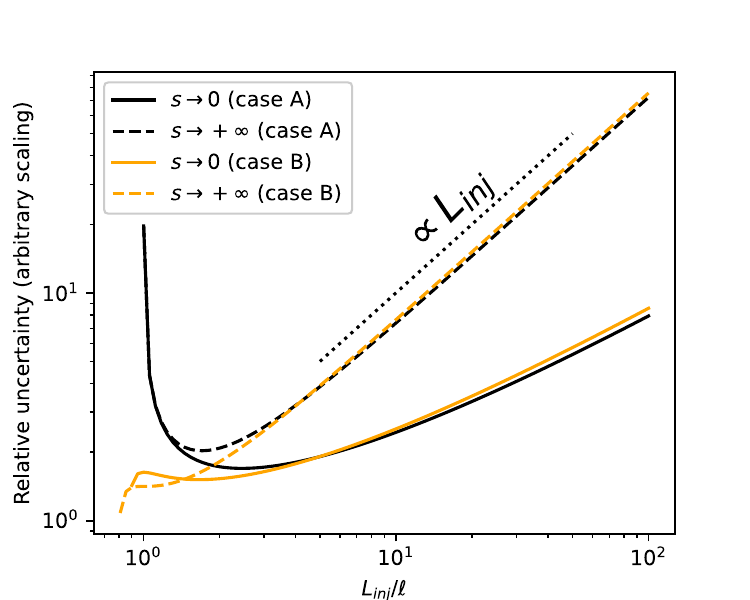}
    \caption{Approximate relative uncertainty (sample variance only) in the two cases, A and B, and for the limiting regimes of small and large separations, $\sepa$. In this figure, we fix the bin size, $\ell$, the area, $\fovarea$, and the slope, $\alpha$, letting only the injection scale vary (x-axis, in units of the bin size, $\ell$).}
    \label{fig:appendix_approx_error_linj}
\end{figure}

\begin{figure}
    \centering
                \includegraphics[width=\linewidth]{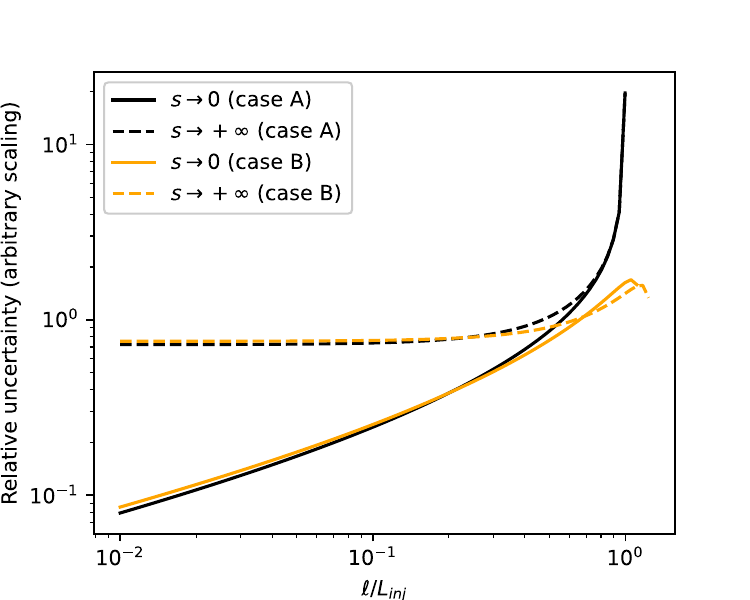}
    \caption{Approximate relative uncertainty (sample variance only) in the two cases, A and B, and for the limiting regimes of small and large separations, $\sepa$. In this figure we fix the injection scale, $L_{inj}$, the area, $\fovarea$, and the slope, $\alpha$, letting only the bin size, $\ell,$ vary (x-axis, in units of $L_{inj}$).}
    \label{fig:appendix_approx_error_binsize}
\end{figure}

In all four limiting cases, it clearly appears that relative uncertainties scale with the ratio $L_{inj}/\sqrt{\fovarea}$ at fixed $r=\ell/L_{inj}$. Larger regions of analysis that sample multiple (projected) injection lengths are less prone to sample variance, as one might expect.
Figure~\ref{fig:appendix_approx_error_linj} is obtained by fixing the bin size $\ell$ and letting only the injection scale vary. The relative uncertainty scales linearly, with $L_{inj}$ in the regime of large separations, if $L_{inj}$ is sufficiently larger than the bin size.
The dependence with the bin size $\ell$ and the slope of turbulence, $\alpha,$ through the functional $\mathcal{G}$ is less intuitive. Generally, such larger bin sizes are known to induce larger relative variance; whereas steeper power spectra also induce larger relative sample variance. Figure~\ref{fig:appendix_approx_error_binsize} shows the dependency of relative uncertainties upon the bin size $\ell$, all other quantities held fixed. In the regime of large separations, the relative uncertainties does not depend much on the bin size, provided bins are small enough.

\subsection{Exact calculations}

\begin{figure}
    \centering
                \includegraphics[width=\linewidth]{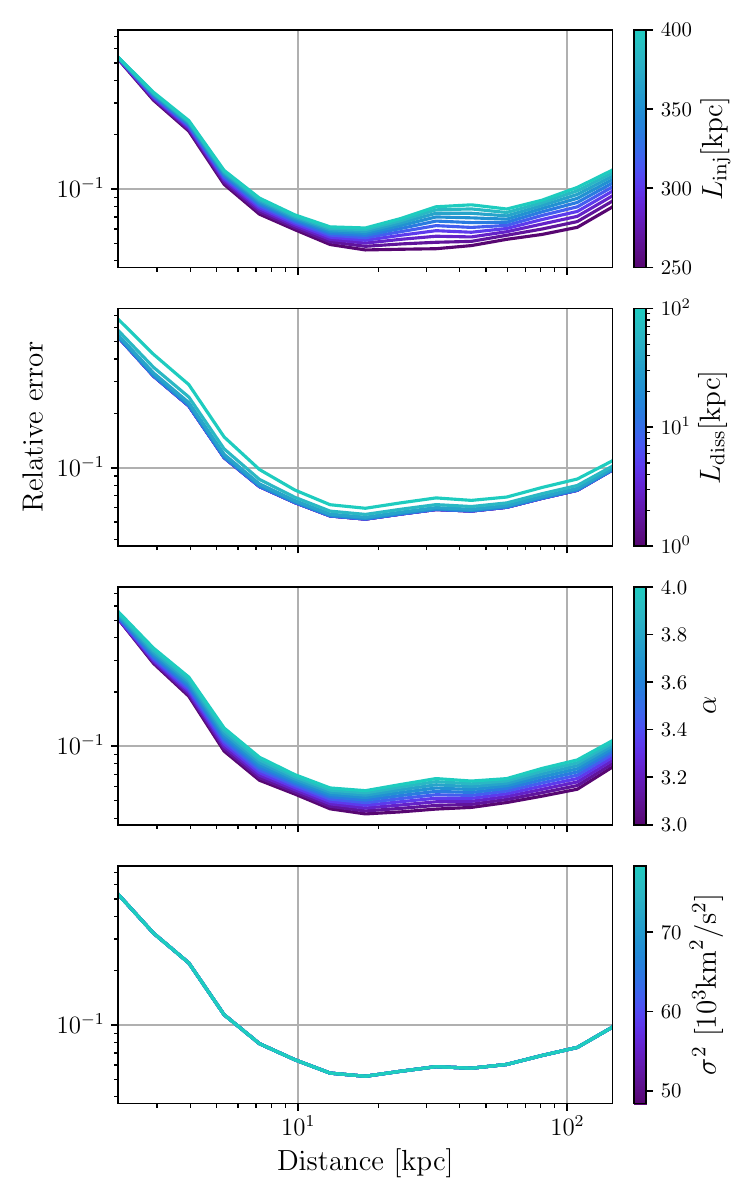}
    \caption{Exact relative error computation of the structure functions for the mix exposure observation, as a function of separation. Only the sample variance is included in the error budget.}
    \label{fig:appendix_varSFexact}
\end{figure}

\indent The integrals in Eqs.~\ref{eq:appendix_sf_mean} and~\ref{eq:appendix_sf_var} (and in Sect~\ref{s:sferrors}) for the sample variance computation can be computed exactly, namely,~without approximations. This is the approach used in the core of this work. In Fig.~\ref{fig:appendix_varSFexact}, we show the relative error obtained for different values of the turbulent parameters as a function of pair separation, $\sepa$. We employ the same method as described in Sect.~\ref{s:sferrors} to associate each value of $\sepa$ with a neighboring, best-matching Xbeta emissivity model, with its associated geometry (i.e.,~bin size and total area).
As expected, the resulting relative error is independent of the norm of the power spectrum (bottom panel). Similarly as for the structure function, dissipation scales smaller than the bin size do not affect the relative error and even above this size, they have a limited influence. The injection scale has a significant impact on the relative error at large separations, with larger injection scales resulting in larger errors.

\end{document}